\documentclass[a4paper,11pt]{article}
\pdfoutput=1 % if your are submitting a pdflatex (i.e. if you have
\usepackage{jinstpub} % for details on the use of the package, please
\usepackage{amsmath}
\usepackage{amsfonts}
\usepackage{enumitem}
\usepackage{ulem}

\usepackage[dvipsnames]{xcolor}

\usepackage{lineno}

\title{Design and implementation of the new scintillation light detection system of
ICARUS T600}

\collaboration[c]{for the ICARUS Collaboration}

\author[a]{B.~Ali-Mohammadzadeh,}
\author[b,c]{M.~Babicz,}
\author[d]{W.~Badgett,} 
\author[d]{L.~Bagby,}
\author[a]{V.~Bellini,}
\author[e]{R.~Benocci,} 
\author[e]{M.~Bonesini,} 
\author[f,g]{A.~Braggiotti,} 
\author[f]{S.~Centro,}
\author[h,d]{A.~Chatterjee,}
\author[i]{A.G.~Cocco,}
\author[l]{M.~Diwan,}
\author[e]{A.~Falcone,} 
\author[f]{C.~Farnese,}
\author[d]{A.~Fava,}
\author[f]{D.~Gibin,}
\author[f]{A.~Guglielmi,}
\author[d]{W.~Ketchum,} 
\author[c]{U.~Kose,}
\author[m]{A.~Menegolli,}
\author[f]{G.~Meng,}
\author[m,d]{C.~Montanari,}
\author[c]{M.~Nessi,}
\author[c,f]{F.~Pietropaolo,}
\author[m]{A.~Rappoldi,}
\author[m,1]{G.L.~Raselli\note{Corresponding author.},}
\author[m]{M.~Rossella,}
\author[c,n,o]{C.~Rubbia,} 
\author[p,c]{P.~Sala,}
\author[m]{A.~Scaramelli,}
\author[c,q]{F.~Sergiampietri,}
\author[l,2]{M.~Spanu\note{now at 
Department of Physics, University of Milano Bicocca and INFN, Milano, Italy},}
\author[d]{D.~Torretta,}
\author[e]{M.~Torti,}
\author[a]{F.~Tortorici,}
\author[f]{F.~Varanini,}
\author[f]{S.~Ventura,}
\author[o]{C.~Vignoli,}
\author[l,3]{A.~Zhang\note{now at Department of Physics and Astronomy, Stony Brook University, Stony Brook NY, USA}}
\author[p]{and A.~Zani}

\affiliation[a]{Department of Physics and Astronomy, University of Catania and INFN, Catania, Italy}
\affiliation[b]{Institute of Nuclear Physics PAN, Krak\'ow, Poland}
\affiliation[c]{CERN, Geneve, Switzerland}
\affiliation[d]{Fermi National Laboratory, Batavia IL, USA}
\affiliation[e]{Department of Physics, University of Milano Bicocca and INFN, Milano, Italy}
\affiliation[f]{Department of Physics and Astronomy, University of Padova and INFN, Padova, Italy}
\affiliation[g]{CNR Padova, Padova, Italy}
\affiliation[h]{Department of Physics and Astronomy, University of Pittsburgh, Pittsburgh PA, USA}
\affiliation[i]{Department of Physics, University of Napoli and INFN, Napoli, Italy}
\affiliation[l]{Brookhaven National Laboratory, Brookhaven NY, USA}
\affiliation[m]{Department of Physics, University of Pavia and INFN, Pavia, Italy}
\affiliation[n]{Gran Sasso Science Institute, L'Aquila, Italy}
\affiliation[o]{INFN, Laboratori Nazionali del Gran Sasso, Assergi, Italy}
\affiliation[p]{Department of Physics and INFN, University of Milano, Milano, Italy}
\affiliation[q]{INAF Torino, Torino, Italy}
\emailAdd{gianluca.raselli@pv.infn.it}

\abstract{
ICARUS T600 is the far detector of the Short Baseline Neutrino program at Fermilab (USA), which foresees
three Liquid Argon Time Projection Chambers along the Booster Neutrino Beam line
to search for LSND-like sterile neutrino signal.
The T600 detector underwent a significant overhauling process at CERN, introducing
new technological developments while maintaining the already achieved
performances.
The realization of a new liquid argon scintillation
light detection system is a primary task of the detector overhaul. As the detector will be subject
to a huge flux of cosmic rays, the light detection system should allow the 3D reconstruction of
events contributing to the identification of neutrino interactions in the beam spill gate. 
The design and implementation of the new scintillation light detection system of
ICARUS T600 is described.}


\begin{document}

\maketitle
%%%\linenumbers

\section{Introduction}

The ICARUS T600 detector is the largest Liquid Argon Time Projection Chamber (LAr-TPC)
ever operated on a neutrino beam for oscillation studies. 
It took data from 2010 to 2013 in the INFN Gran Sasso Laboratory (Italy), both with atmospheric neutrinos and with the
CERN Neutrinos to Gran Sasso (CNGS) beam. 
After an intense refurbishing operation, carried out at CERN
in the framework of the Neutrino Platform
activities (WA104/NP01), the entire apparatus was moved to Fermilab (IL, USA), where it will operate as far detector 
of the Short Baseline Neutrino (SBN) program~\cite{Antonello:2015lea}: three liquid argon detectors, placed along the Booster Neutrino Beam (BNB) line and operating at shallow depth, will investigate the possible presence of sterile neutrino states.

The realization of a new light detection system, sensitive to the photons produced by
the LAr scintillation, is a fundamental feature 
for the T600 operation at shallow depth. 
A threshold of 100 MeV of deposited energy, a time resolution of the order of $\approx 1$~ns and a high
granularity are required to effectively identify the events associated to the neutrino beam and handle
the expected huge cosmic background.
The T600 scintillation light detection system was significantly upgraded
at CERN from summer 2015 to summer 2017, after preliminary studies
based on simulations and laboratory tests, devoted to optimizing the
performance of the apparatus.

This paper presents the main characteristics and the realization of the new 
scintillation
light detection system of ICARUS T600.
Section~\ref{detector} describes the ICARUS T600 detector. 
%The main features of the light detection system are presented in section~\ref{light}. 
The design and optimization of the new light detection system is presented in Section~\ref{design}. 
Section~\ref{system}
reports the characteristics of the various components and their implementation.
%, illustrates the realization of the system and
Finally some preliminary system tests, carried out after the transportation to Fermilab
followed by the installation of the apparatus in the far detector building,
are presented in Section~\ref{tests}.

\section{The ICARUS T600 detector}
\label{detector}

The ICARUS T600 detector is made of two identical cryostats, 
filled with about 760~t of ultra-pure liquid argon~\cite{Amerio:2004ze}.
Each cryostat houses two TPCs with 1.5~m maximum drift path, sharing a common
central cathode made of punched stainless-steel panels.
The cathode plane and field cage electrodes, composed by stainless-steel tubes, 
generate an ideally uniform electric field $E=500$~V/cm.

Charged particles interacting in liquid argon produce both scintillation light
and ionization electrons. Electrons are drifted by the electric field
to the anode, made of three parallel wire planes.
A total of 53248 wires are deployed, with 3~mm pitch, oriented on each plane at a different angle 
($0^\circ$, $\pm 60^\circ$) with respect to the horizontal direction. 
By appropriate voltage biasing, the first two wire planes
record signals in a non-destructive way, while the ionization charge is collected and measured on the
last plane. 
The electronics was designed to allow continuous read-out, digitization
and independent waveform recording of signals from each wire of the TPC,
with 400~ns sampling time and 12-bit dynamic range~\cite{Bagby:2018fkj}.
The information of the ionization track occurrence time, combined with the electron drift velocity 
($ v \approx 1.6$~mm/$\mu$s at $E=500$~V/cm)
provides the event coordinate in the drift direction. The composition of the three views
from the TPC wires yields the track projection on the anode plane. This information allows obtaining
a full 3D reconstruction of the tracks, with a spatial resolution of about 1~mm$^3$~\cite{Antonello:2012hu}.

The precise information of the event occurrence time
is given by the LAr scintillation light
which permits 
the generation of a light-based trigger signal and a preliminary 
identification of event topology for fast selection purposes~\cite{Antonello_2014}.
The light information is a fundamental feature for the identification of 
signals related to the neutrino beam induced events. 
This requires a high performance light detection system as described in the following sections.

\section{Design and optimization of the scintillation light detection system}
\label{design}

\subsection{Scintillation light emission in liquid argon}
\label{light}

Scintillation light emission in LAr is due to the radiative decay of excimer molecules Ar$_2^*$ produced by ionizing particles, releasing monochromatic VUV photons ($\lambda \approx 128$~nm) in transitions from the lowest excited molecular state to the dissociative ground state. 
The emitted light is characterized by a fast
($\tau \approx 6$~ns) and a slow ($\tau \approx 1.5$~$\mu$s) decay components. 
%A fast ($\tau \approx 6$~ns decay time) and a slow ($\tau \approx 1.5$~$\mu$s) components are emitted. 
Their relative intensity depends on $dE/dx$, ranging from 1:3 for minimum ionizing particles, up to 3:1 for alpha particles. %, allowing in some particular cases a way to perform a particle discrimination. 
This isotropic light signal propagates with negligible attenuation throughout each TPC volume. Indeed, LAr is fully transparent to its own scintillation light, with measured attenuation length in excess of several tens of meters and Rayleigh-scattering length of about 1~m~\cite{Babicz:2020den}. %~\cite{Babicz:2018gqv}. 
Because of their short wavelength the scintillation photons are absorbed by all detector materials without reflection, leaving time and amplitude information unaffected during the photon path to the light detectors. 
%Since photons from scintillation light propagate almost unaffected from the production point to the light detectors, they keep information about the time of generation, i.e. the ionizing particle interaction time and the time evolution of the interaction event in LAr. 

\subsection{Scintillation light detection in ICARUS T600}

A scintillation light detection system based on 74 ETL9357FLA ($8''$ diameter) PMTs mounted behind the wire chambers was adopted
in the T600 detector for the LNGS run~\cite{Ankowski:2006xx}. The sand-blasted glass window of each device was coated with about 200~$\mu$g/cm$^2$  of Tetraphenyl Butadiene (TPB), to convert the VUV photons to visible light. ICARUS at Fermilab will take data at shallow depth, facing more challenging experimental conditions than at LNGS.  The light detection system will complement the 3D track reconstruction 
performed with the use of the TPC wires, thus contributing to identify neutrino interactions occurring in the BNB spill gate structure
and rejecting the expected $\approx 10$~kHz cosmic background. This new environment requires a number of improvements, namely the adoption 
of a PMT model with better performances, an improvement of the sensitivity down to 100~MeV, a time resolution $\mathcal{O}$(1~ns) and an increase 
of the light detection granularity. This last requirement is needed to localize the track associated with every light pulse along the
$\approx 20$~m of the longitudinal detection direction, with accuracy better than 1~m, namely shorter than the expected 
average spacing between cosmic muons in each TPC image.  In this way, the light detection system would be able to unambiguously provide
the absolute timing for each track, and to identify, among the several tracks in the LAr TPC image, the event
in coincidence with the neutrino beam spill. 

The adoption of large area PMTs coated with TPB was considered the best solution for the light detection system upgrade.
The use of alternative devices, such as SiPM detectors, was considered not mature enough for applications in large volume LAr-TPC because of their small
sensitive surface.

\subsection{Optimization by Monte Carlo simulation}
\label{montecarlo}

%{\color{BlueViolet} 

Dedicated Monte Carlo simulations were realized to design and optimize the light detection system for the refurbishing of the T600 detector.

Initially, the focus was put on the geometrical properties
%The first one to be developed was based on the geometrical properties 
of the propagation of the VUV scintillation light in ICARUS, with some simplifications on the features of the topology of the considered class of events~\cite{Falcone:tesi}:

\begin{enumerate}
\setlength\itemsep{0pt}
\item electromagnetic (e.m.) showers, mimicking Neutral Current (NC) and $\nu_e$ Charged Current (CC) interactions from BNB;
\item single crossing cosmic muons, which represent the most abundant source of background;
\item muons generated from $\nu_\mu$ CC interactions.
\end{enumerate}

Fine details of physical events, such as e.m. showers shape or particle multiple scattering,
were found to be less important than the spatial resolution achieved with $8''$ diameter devices spaced by $\mathcal{O}$(1 m). 
%Fine details of physical events, such as e.m. showers shape or particle multiple scattering, were not taken into account because they are less relevant with respect to the spatial resolution that can be achieved with $8''$ diameter devices spaced by $\mathcal{O}$(1 m). 
Muons were schematized as straight lines, while e.m events as clusters of points with 1~MeV deposited energy each. 
Showers energy spanned from 100 MeV to 1~GeV, to cover all the expected energy range in the SBN configuration.
Muons generated from $\nu_\mu$ CC interactions were a superposition of the other two event topologies.
From each point along the simulated track, the proper number of photons was generated isotropically; due to the short wavelength, LAr scintillation light is absorbed by all the detector material, so no reflection was assumed.
%A Rayleigh scattering length of 90~cm was instead considered, which is confirmed by recent studies~\cite{Babicz:2018gqv}.
A Rayleigh scattering length of 90~cm was considered. %which is confirmed by recent studies~\cite{Babicz:2018gqv}.
A 5\% overall Quantum Efficiency (QE) was conservatively assumed for the PMTs,
%: this values agrees with the experimental measurements [1] 
which includes wavelength shifting conversion efficiency and geometrical factor: about 50\% of the light is loss during conversion.
This QE value was adopted following the working hypothesis  of using ETL9357FLA PMTs coated with TPB 
by means of a spraying technique, as realized for the LNGS run~\cite{Ankowski:2006xx}. %    ~cite{Amerio:2004ze}.}
An error of $\pm 1$~ns on the arrival time and a $\pm$10\% uncertainty on the number of collected photons were assumed, to take into account 
the PMT response uncertainties according to experimental measurements results~\cite{Falcone:tesi,Falcone:2017zhz}. 
%the foreseen instrumental error.

Different PMT positioning layouts with $8''$ and $5''$ diameter PMTs were considered to study the performance both for cosmic muons and for e.m. showers in the T600 detector. Configurations with different numbers of PMTs were also considered, starting from 27 devices ($8''$ PMTs) up to 210 devices ($5''$ PMTs) per TPC. %~\cite{Falcone:2017zhz}. 
The pattern of each layout was constrained by the existing mechanical structure of the T600: the requirement was not to change it, exploiting the free space already available in this structure.

\begin{figure}[!ht]
\centering
\includegraphics[width=0.45\columnwidth]{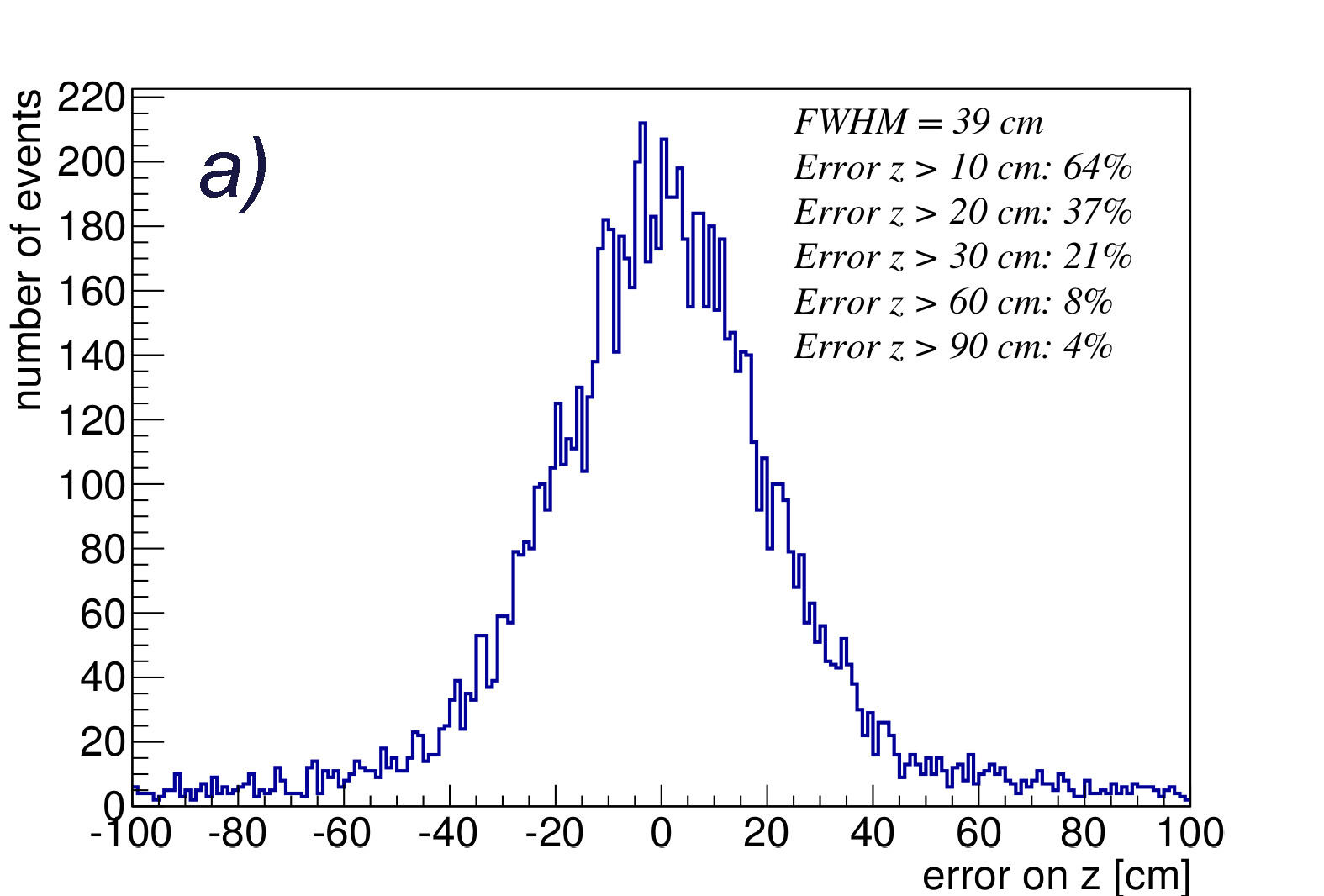}
\includegraphics[width=0.45\columnwidth]{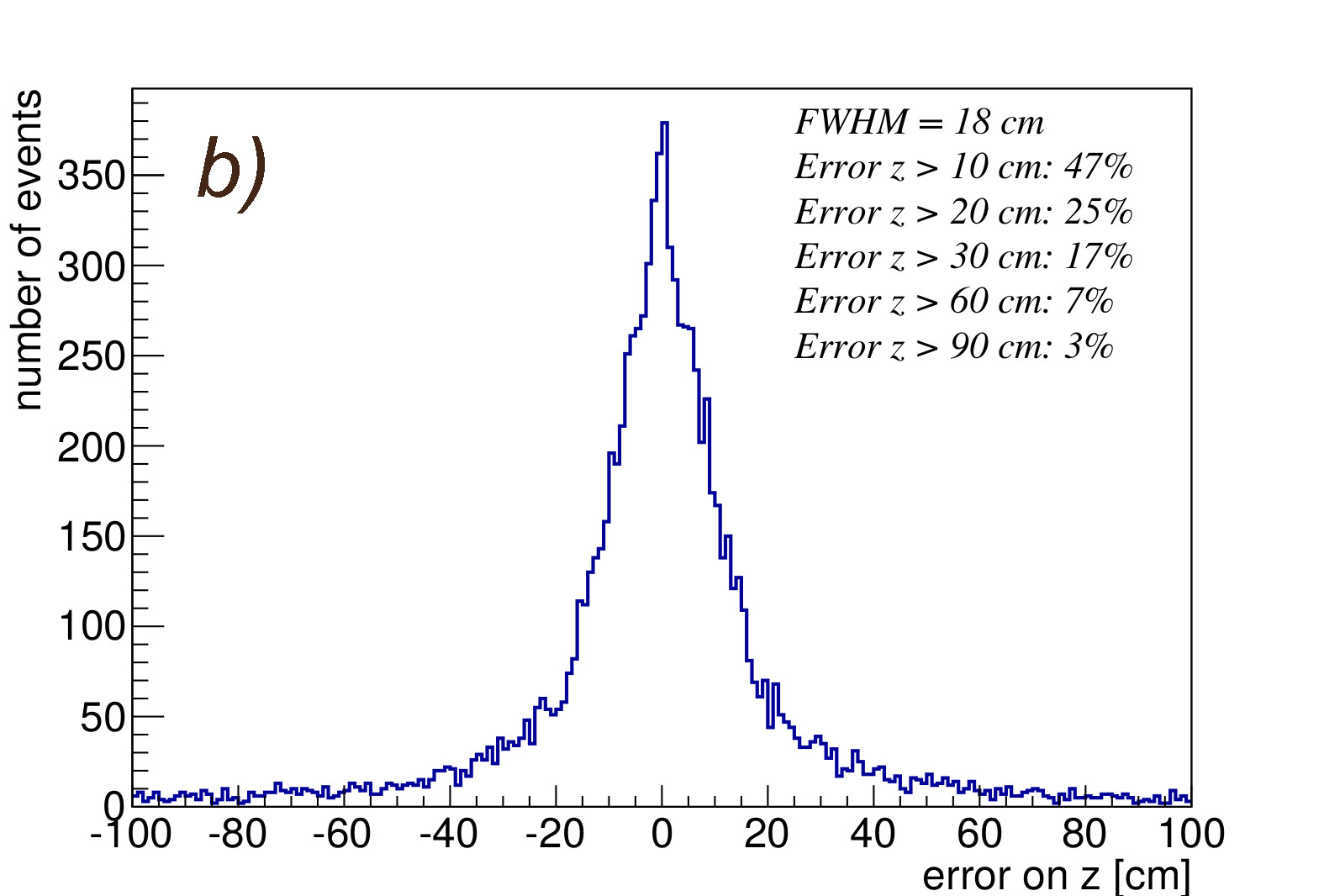}\\
\includegraphics[width=0.45\columnwidth]{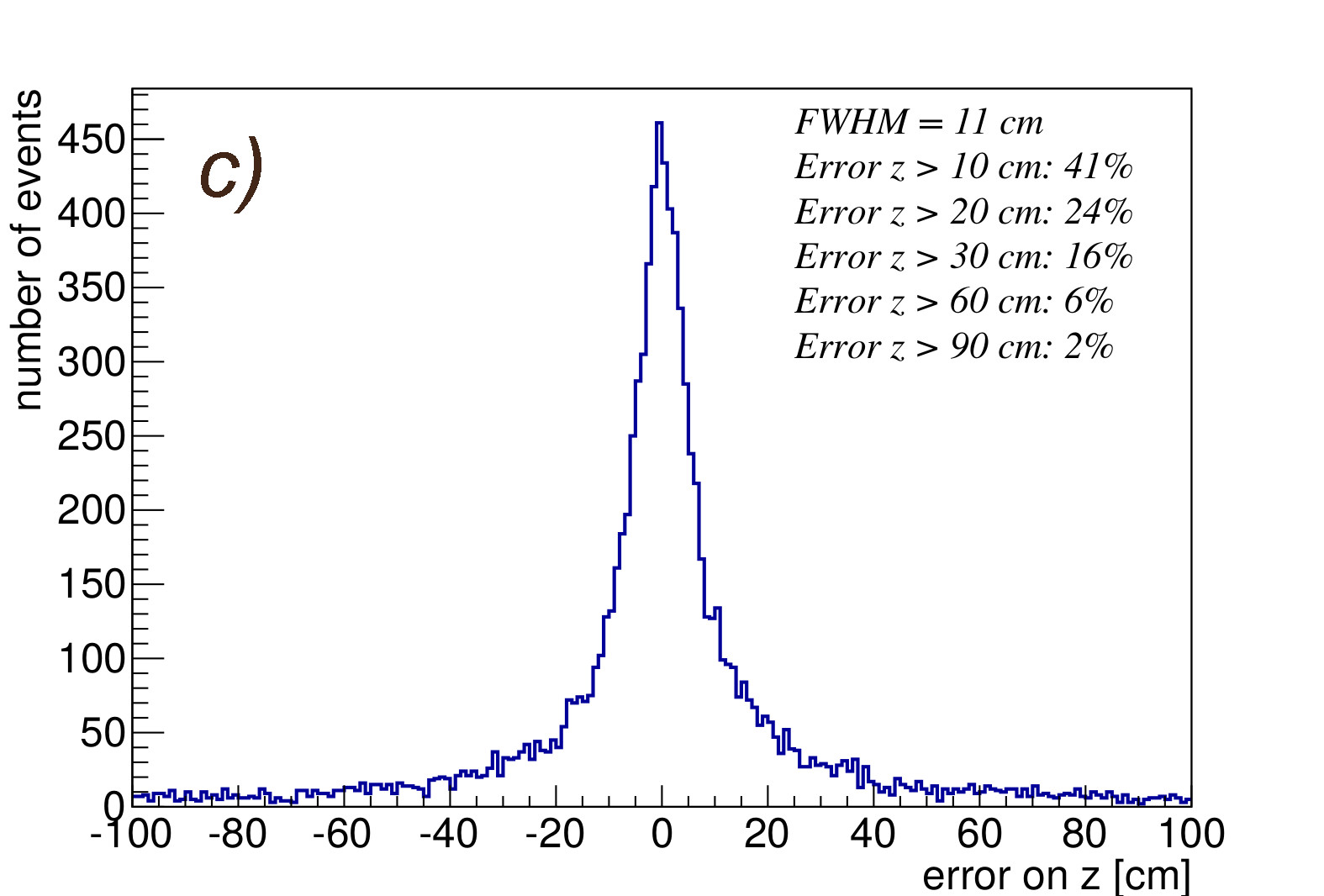}
\includegraphics[width=0.45\columnwidth]{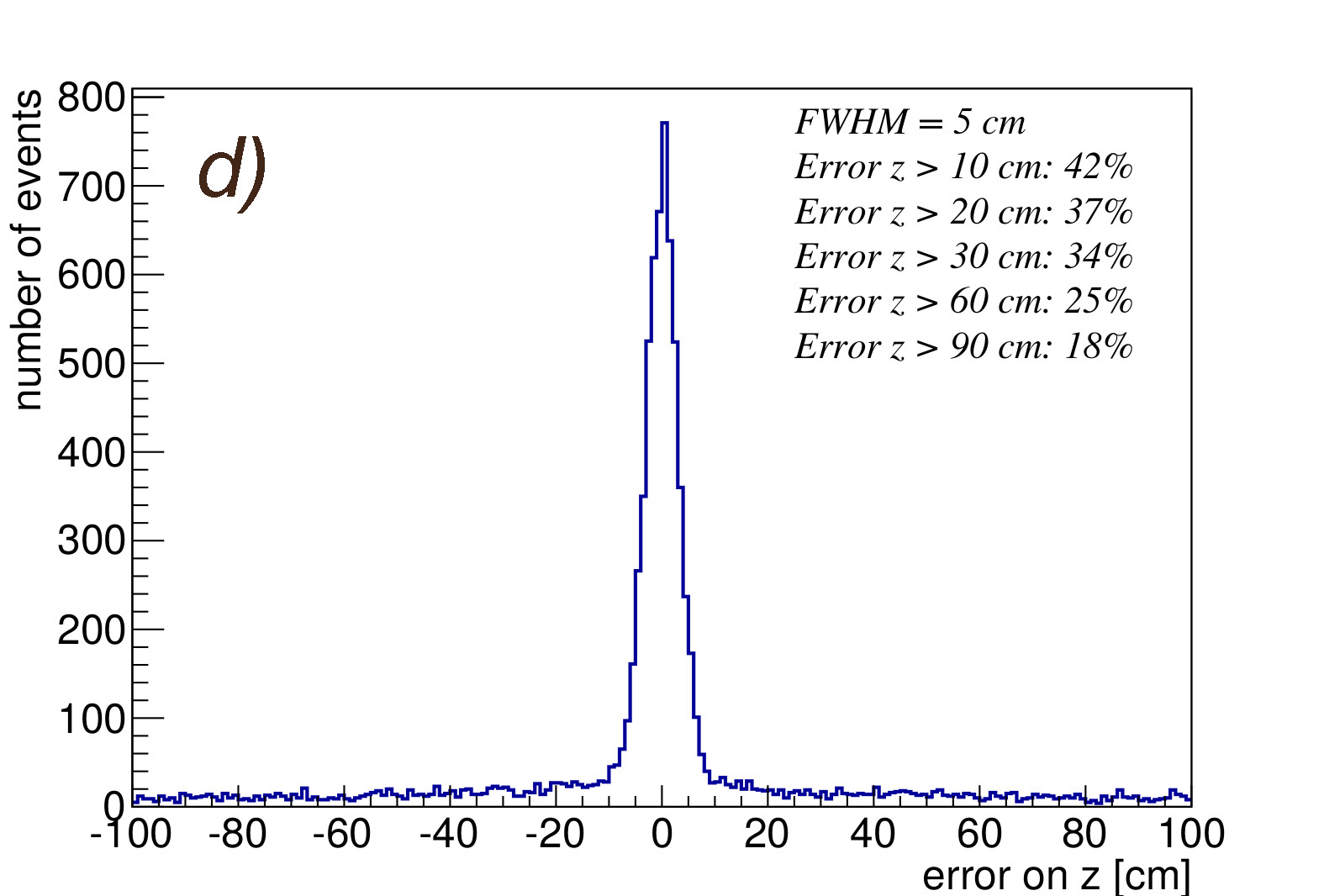}
\caption{Evaluation of the precision on the localization of the actual interaction position along the beam direction z for e.m. showers as determined with the custom simulation. The considered layouts are: {\it a)}\/ 27 ($8''$) PMTs;   {\it b)}\/ 90 ($8''$) PMTs; {\it c)}\/ 210 ($5''$) PMTs.
Figure  {\it d)}\/  shows the result 
obtained with the layout with 90 ($8''$) PMTs
by considering only those devices with a signal above a threshold of 10~phe.
The FWHM and the percentage of events localized with an error greater than 10, 20, 30 60 and 90 cm are indicated in each figure.}
\label{fig_localize}
\end{figure}

For what concerns the event position reconstruction, simulations were carried out to evaluate the capabilities of the different configurations to localize the e.m. showers, mainly along the 18~m length on the beam direction (z axis). The error on the event position reconstruction was calculated as the difference between the actual geometrical center of the event and the one derived from the average on the PMT coordinates, weighted on the light collected by each PMT. The best results were obtained by the set of geometries with the highest numbers of PMTs, as shown in figure~\ref{fig_localize} for
{\it a)}\/ 27 ($8''$) PMTs,   {\it b)}\/ 90 ($8''$) PMTs, and  {\it c)}\/ 210 ($5''$) PMTs.
%which was obtained with the custom simulation. 
Anyway the difference among them is not significant and
performance improvements can be obtained just by refining the reconstruction algorithm. For example, just considering in the average on the PMT position only those devices with a signal above a threshold of 10~phe, as shown in figure~\ref{fig_localize} $d)$, the 90 ($8''$) PMTs configuration shows a localization capability which is better than the one obtained with the 210 ($5''$) PMTs configuration. 

A more detailed study of the PMT system performance was then carried out within LArsoft, which is a framework supporting a shared base of physics software across LAr-TPC experiments. In particular, the ICARUS T600 detector description and the particle generation and propagation in the ICARUS volume are determined within Geant4 software which is implemented in LArsoft. The ICARUS T600 inner detectors main components (wires, PMTs, cathode, field cage) are faithfully reproduced. Both single BNB $\nu_\mu$ and $\nu_e$ interactions and cosmic ray samples were generated inside the active volume of one of the two modules of ICARUS.
LArsoft framework covers basic components of the real scintillation detector system and includes all relevant physical processes. With the complete simulation, individual physical factors that can affect the performance of the detector system, such as detector geometry, surface finishing, decay time and scintillation yield of scintillator as well as 
the actual response of PMTs, presented in Sections~\ref{PMT} and \ref{evaporation}, and front-end electronics, are taken into account \cite{light_dgg}.

\begin{figure}[!ht]
\centering
\includegraphics[width=0.49\columnwidth]{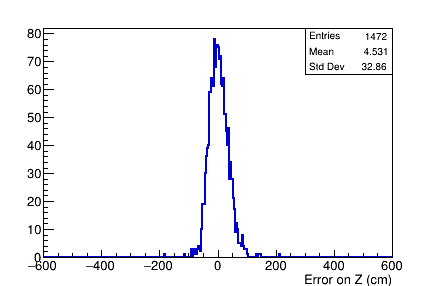}
\includegraphics[width=0.49\columnwidth]{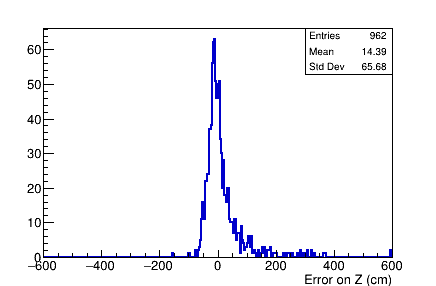}
\caption{LArsoft evaluation of the precision on the localization of the actual neutrino vertex position, along the beam direction z, for $\nu_e$ 
({\it Left}\/) and $\nu_\mu$ ({\it Right}\/) for the layout with 90 ($8''$) PMTs: a precision of about 30 cm (70 cm) for $\nu_e$CC ($\nu_\mu$CC) is obtained.}
\label{bary_light}
\end{figure}

The impact of the layout with 90 ($8''$) PMTs for each TPC in terms of neutrino vertex localization, as obtained from LArsoft simulation of BNB $\nu_e$CC and $\nu_\mu$CC events, is illustrated in figure~\ref{bary_light}: provided the timing information from all PMTs is available, an accuracy of less than 15 cm and a precision of about 30 cm (70 cm) for $\nu_e$CC ($\nu_\mu$CC) was obtained by estimating the neutrino vertex position from the light barycenter of the first three hit PMTs.

\begin{figure}[!ht]
\centering
\includegraphics[width=0.8\columnwidth]{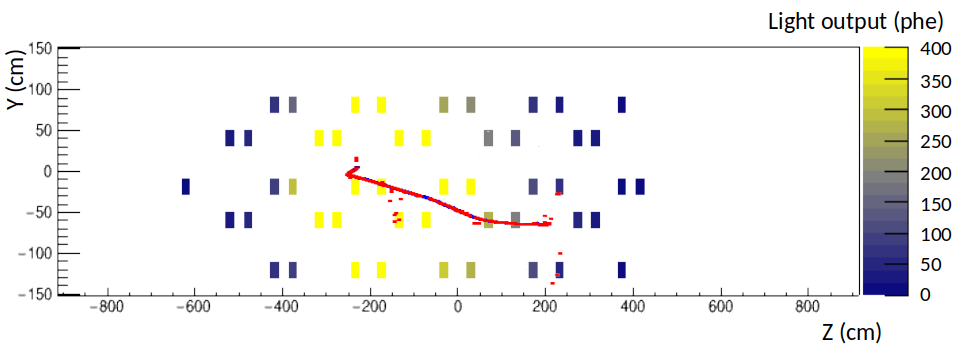}
\includegraphics[width=0.8\columnwidth]{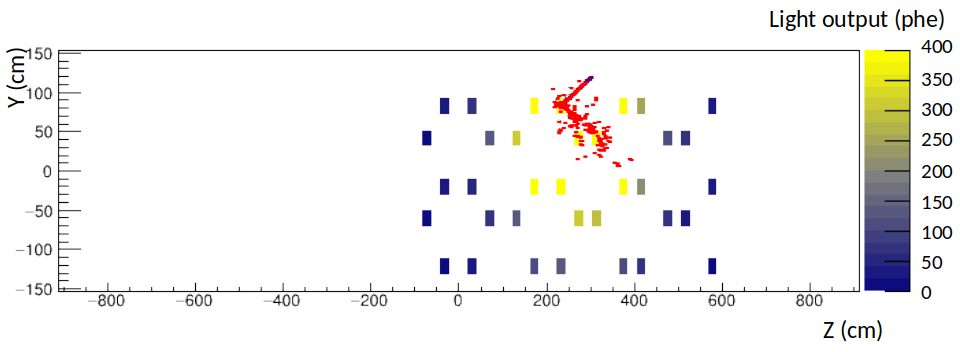}
\caption{Examples of neutrino events simulated with LArsoft and reconstructed from the TPC wires superimposed with the color map of PMTs having a light signal exceeding 10 photo-electrons. Top: $\nu_\mu$CC, 1.2 GeV deposited energy. Bottom: $\nu_e$CC 0.9 GeV deposited energy.}
\label{pmt_map}
\end{figure}

Finally, this analysis opens 
%Finally, an analysis on
the possibility to directly match the event position as determined by the analysis of the charge with light information coming from PMTs. This could help in developing a quick first level event tagging using the light signals. 
As an example of the localization capability,
%As an example of the possibilities in localizing events by the scintillation light, 
two neutrino events reconstructed from the TPC wires are shown superimposed with the map of PMTs in in figure~\ref{pmt_map}.

%}

The described MC simulations led to select the 90 ($8''$) PMTs layout for
installation on the ICARUS T600 TPCs.
%Monte Carlo simulation results presented in this section deemed adequate a layout with 90 ($8''$) PMTs for each TPC.
This layout implies the use of 360 PMTs with $8''$ diameter, corresponding to a coverage of 5\% of the wire plane surface.
The estimation of the number of photo-electrons collected per MeV of deposited energy in a single TPC gives an average of about 15~phe/MeV (9~phe/MeV for events close to the cathode). The possibility to adopt the scintillation light for triggering and timing purposes with events down to 100~MeV 
%using fairly high signal thresholds and high multiplicity on PMTs 
is then assured in the whole TPC volume.

\section{Hardware implementation}
\label{system}

%Monte Carlo simulation results deemed adequate a layout with 90 ($8''$) PMTs for each TPC.
%The adopted layout, shown in  figure~\ref{fig_deploy}, implies the use of 360 PMTs, corresponding to a coverage area of 5\% of the wire plane surface.
%The estimation of the number of photo-electrons collected per MeV of deposited energy in a single TPC gives an average of about 15~phe/MeV (9~phe/MeV for events close to the cathode). The possibility to trigger low energy events (100~MeV) with fairly high threshold and multiplicity on PMTs is then assured in the whole TPC volume.

\begin{figure}[!ht]
\centering
\includegraphics[width=1.0\columnwidth]{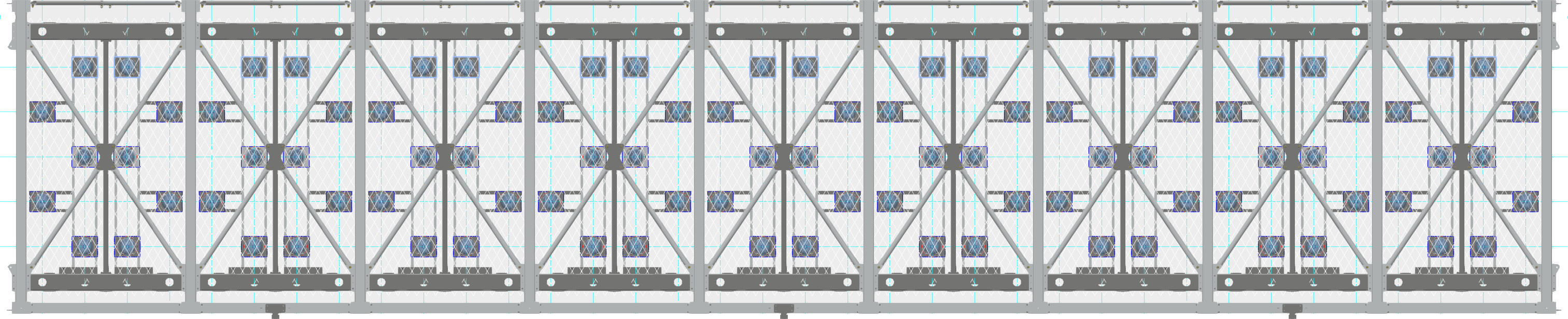}\\
\vspace{24pt}
\includegraphics[width=0.75\columnwidth]{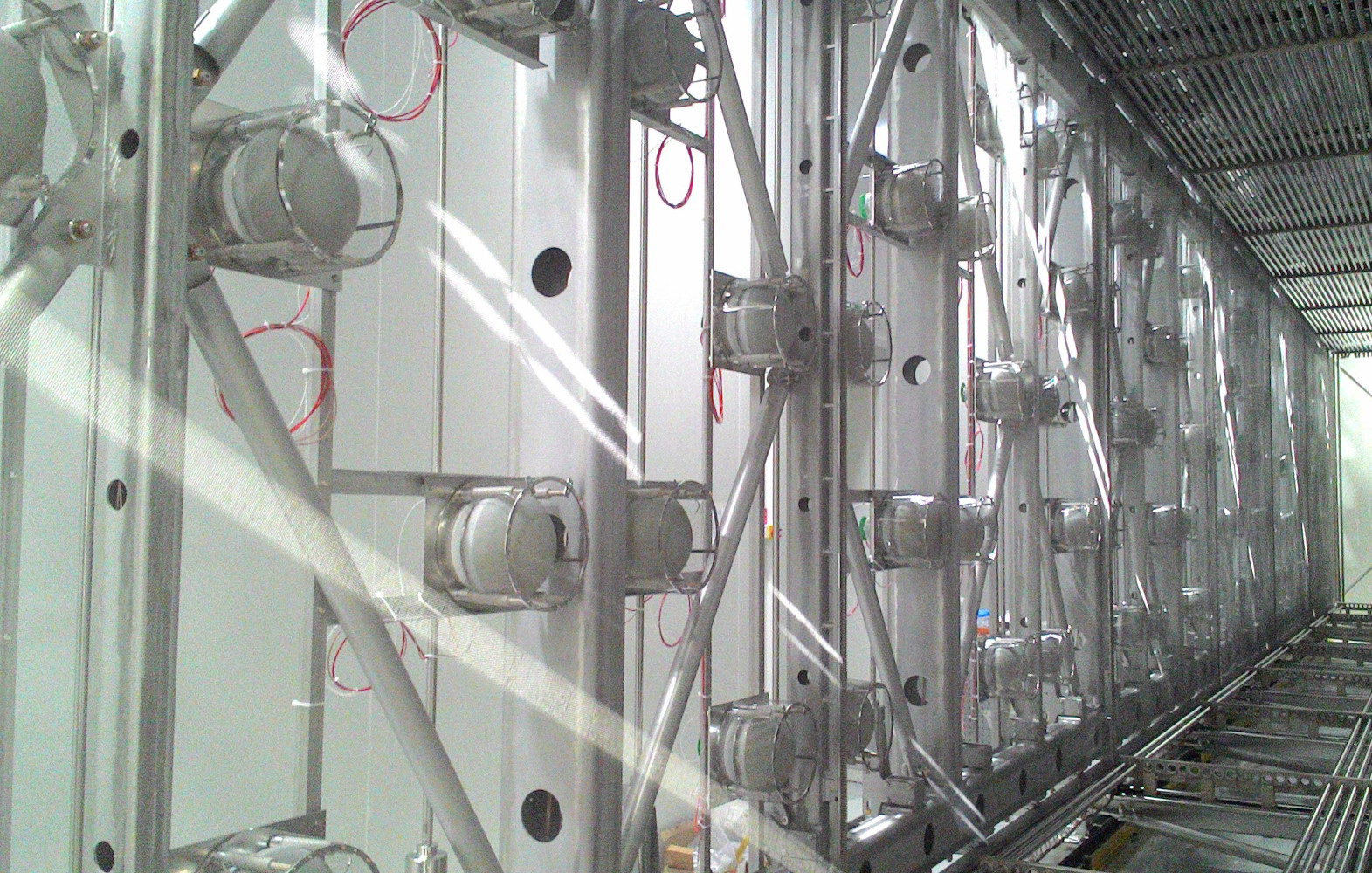}
\caption{Scheme of the adopted geometry with 90 PMTs behind each wire plane and picture of the actual configuration.}
\label{fig_deploy}
\end{figure}

%The adopted PMT layout, shown in  figure~\ref{fig_deploy}, implies the use of 360 devices with $8''$ diameter, and corresponds to a coverage area of 5\% of the wire plane surface.
%The realization of this layout 
%The new light detection system of ICARUS T600 
The realized PMT layout, shown in  figure~\ref{fig_deploy},
features 360 total Hamamatsu R5912-MOD PMTs
%Its realization of the PMT layout, shown in  figure~\ref{fig_deploy},
%is based on a total of 360 Hamamatsu R5912-MOD PMTs 
deployed in groups of 90 devices behind
each wire chambers.
Since the PMT glass windows is not transparent to the scintillation light produced in liquid argon,
each unit was coated with a proper wavelength shifter re-emitting in the visible. 
The PMTs were installed using dedicated mechanical supports. 
A laser calibration system permits the timing calibration of the single units. 

In this section the ICARUS T600 scintillation light detection system
is described, presenting the main characteristics of its different components.

\subsection{Photomultiplier tubes}
\label{PMT}

%%%%%The Hamamatsu R5912-MOD, {\color{BlueViolet} whose main features are summarized in table~\ref{table1},}
%%%%%is a 10-dynode-stage PMT, featuring an $8''$ hemispherical sand-blasted glass window with bialkali photocathode
%%%%%on platinum undercoating, which guarantees high performance
%%%%%at cryogenic temperatures (see figure~\ref{fig_pmt}).
%{\color{BlueViolet} 
%This PMT is derived from the parent model R5912.} The bialkali photocathode is deposited
%on a Pt underlayer in order to extend the range of the operating temperature down to about -200~$^\circ$C.

%%%%%The model was selected as a result of experimental tests carried out on different 
%%%%%samples provided by different producers, such as Hamamatsu and ETL~\cite{Agnes:2014rda,Falcone:2015pia,Falcone:2016xgw}. %,Babicz:2018fpf}.
%{\color{BlueViolet} 

In order to identify the most suitable model to the requirements of the light detection system of ICARUS T600, 
a test campaign was carried out on different PMT samples manufactured 
by different producers, such as Hamamatsu and ETL~\cite{Agnes:2014rda,Falcone:2015pia,Falcone:2016xgw,Babicz:2019pll}. %,Babicz:2018fpf}.
All the PMTs taken into consideration feature an $8''$ hemispherical glass window with bialkali photocathode
on platinum undercoating, to guarantee high performance at cryogenic temperatures.
The evaluation of their conformity was based on the following considerations:

\begin{itemize}[leftmargin=*]

\item[-] 
The scintillation light detection system should guarantee a good sensitivity to ionizing interactions in LAr down to an energy deposition of 100~MeV. 
To this end the quantum efficiency and its uniformity over the sensitive surface of PMTs have a strong impact on the global efficiency of the system. The effective values of these parameters resulting from actual measurements on PMT prototypes are considered distinguishing features for the model selection. 

\item[-] The dynamics of the scintillation light detection system should permit the recording of the scintillation light fast component pulses and, at the same time, of single photons arriving from the slow component de-excitation. In addition it has to cope with the expected wide variation of light
intensity which depends on the deposited energy and on the  geometry of interactions inside the LAr volume.
%The recording of the whole PMT signal shape is based on fast waveform digitizers with 2~V input range and 0.122~mV resolution (see section~\ref{electronics}). 
Taking into account a standard electronics for PMT signal recording  without pulse amplification, a gain $G \approx 10^7$ at cryogenic temperature is necessary to detect single photons. In addition the PMT dynamics should permit the generation of anode pulses without remarkable saturation
up to hundreds of photoelectrons.
%signal shape characteristics
%The Hamamatsu R5912-MOD signal shape characteristics, 
%i.e a fast leading edge  $\le 4$~ns, 
%of $\approx 4$~ns, a FWHM $\approx 6$~ns, a trailing edge of $\approx 10$~ns 
%typical gains of about $G = 10^7$ for both room and cryogenic temperature, 
%good enough for the detection of single photons.
%In addition Hamamatsu R5912-MOD 
%make this model compliant with the electronics requirements which make use of commercial devices (see section~\ref{electronics}). 
%permits the 
%exploitation of the full dynamic range of electronics.

\item[-] The light collection system should be able to provide unambiguously the absolute
timing of each interaction and %$t_0$ 
identify, among the several tracks in the LAr-TPC image, the
event in coincidence with the neutrino beam spill. 
In order to achieve $\mathcal{O}$(ns) timing resolution, fast PMT pulses are needed.
%In order to achieve 
%an accuracy at nanosecond level in the time reconstruction of events, fast PMT pulses
%are requested. 
Moreover good 
stability of the transit time as a function of temperature and applied voltage, low
time spread and a good uniformity over the PMT sensitive windows surface are required.

%%%%All the tested PMT models presented
%%%%signal shapes with fast leading edge  $\le 4$~ns suitable for timing application.
%The Hamamatsu R5912-MOD signal shape characteristics, 
%i.e a fast leading edge  $\le 4$~ns, 
%of $\approx 4$~ns, a FWHM $\approx 6$~ns, a trailing edge of $\approx 10$~ns 
%%In addition Hamamatsu R5912-MOD resulted in good
%%stability of the transit time as a function of the temperature and the applied voltage, low
%%time spread and good uniformity over the PMT sensitive windows surface ($\pm 1$~ns) allowing
%%a reconstruction incoming events with time accuracy
%%at nanosecond level.

\item[-] 
%The response of a PMT in absence of light is represented by dark pulses and noise due to
%different processes. 
Since a large number of PMTs is used,
the presence of a high dark count rate can affect the detector performance
inducing stochastic coincidences at trigger level. 
From each PMT a single-photo-electron physical background rate of tens of kHz is expected in LAr. 
This rate consists of residual photons produced from the decay of $^{39}$Ar or other radioactive contaminates. 
Therefore an intrinsic dark count rate of a few kHz at cryogenic temperatures is judged acceptable.
%As for each PMT a single photo-electron physical background rates of tens of kHz is expected in LAr,
%mainly consisting of residual photons coming from the decay of $^{39}$Ar or other radioactive contamination,
%an intrinsic dark count rate up to few kHz at cryogenic temperature 
%is judged acceptable.
%This overall single-photon rate can be faced at trigger level with the adoption of a good match between
%discrimination level and coincidence multiplicity. 
On the other hand the absence of bursts, sparkling,
lightening effects or other noise generating pulses above the single photon level is considered mandatory.

\item[-] ICARUS T600 will operate in absence of external magnetic fields, except for the earth's intrinsic field. 
Therefore the main PMT performances should not be degraded by external magnetic fields of the order of gauss 
at different axial orientations.

%\item[-] ICARUS T600 will operate in absence of magnetic fields, except the presence of the natural 
%geomagnetic one. Given this, the main PMT performances %, those related to gain and timing, 
%should remain valid with the application of an external magnetic field of the order of gauss
%with different axial orientation.

\item[-] The adopted devices should withstand low temperatures and the relative high pressure as expected in LAr immersion. 
%All the considered PMT models were object of a series of thermal 
All the considered PMT models were subjected to a series of thermal
shocks to highlight possible mechanical or cracking problems.

\end{itemize}

%At conclusion the Hamamatsu R5912-MOD, whose main features are summarized in table~\ref{table1} and a picture is presented 
%in figure~\ref{fig_pmt}, was selected for
%installation.

%Best results were achieved with Hamamatsu R5912-MOD that was selected for
%installation. The main features and the characteristics resulting from tests are summarized in table~\ref{table1} 
%while a picture is presented 
%in figure~\ref{fig_pmt}.

\begin{figure}[!t]
\centering
\includegraphics[width=0.5\columnwidth]{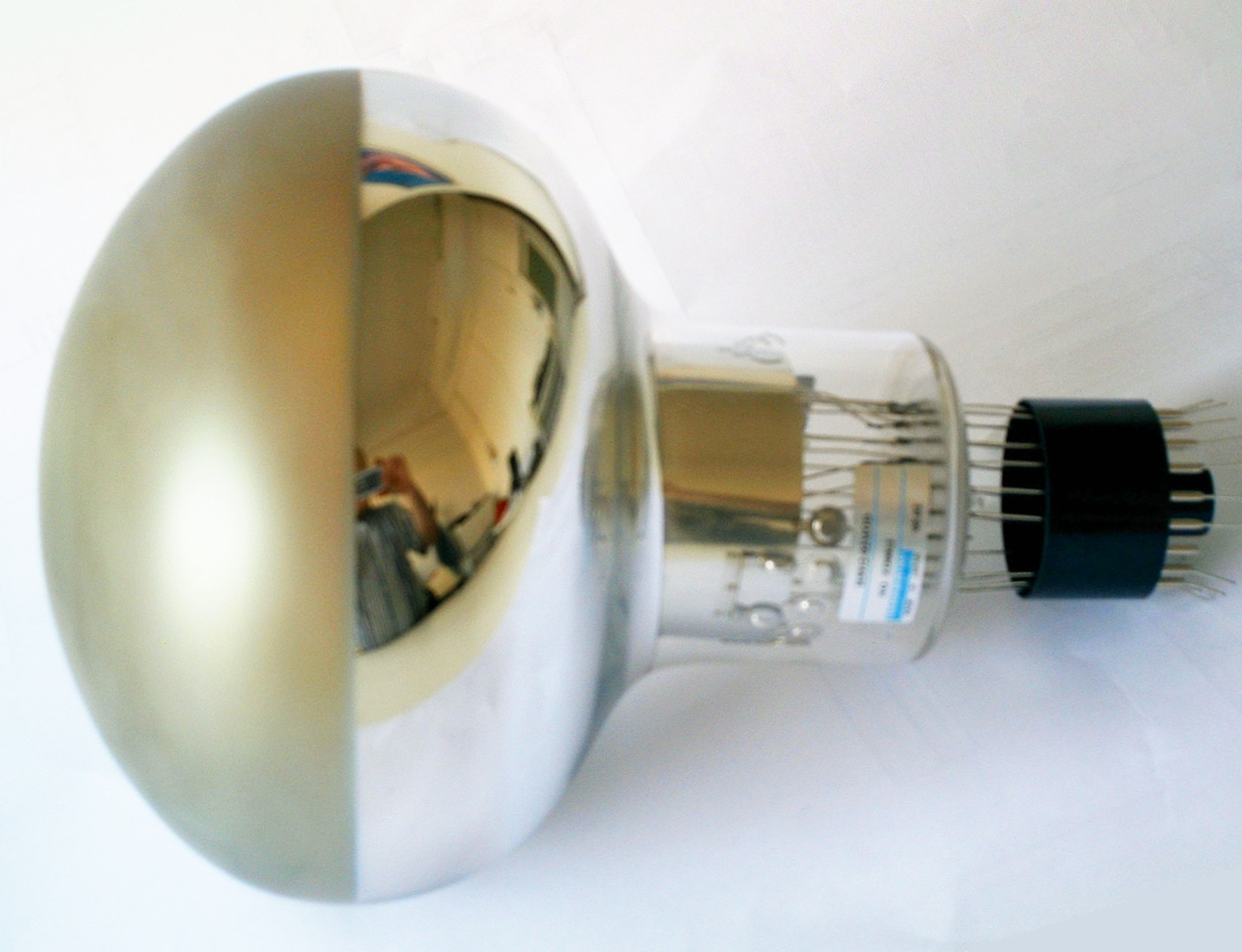}
\caption{The PMT Hamamatsu R5912-MOD.}
\label{fig_pmt}
\end{figure}

Best results were obtained with the Hamamatsu R5912-MOD device (see figure~\ref{fig_pmt}) which was therefore selected for installation.
The main features and the characteristics resulting from the tests are summarized in table~\ref{table1}.

%}

%Results confirm the conformity of this PMT model to the requirements of the
%ICARUS T600 light detection system.
%The main features of the Hamamatsu R5912-MOD are summarized in table~\ref{table1}.

\begin{table}[!t]
\renewcommand{\arraystretch}{1}
\caption{Main acceptance requirements for the Hamamatsu R5912-MOD}
\label{table1}
\centering
\begin{tabular}{ll}
\hline
%\toprule
Spectral Response      & $300\div 650$~nm \\
Window Material        & Borosilicate glass (sand blasted) \\
Photocathode           & Bialkali with Pt under-layer   \\
%Photocathode min. area & 190 mm diameter \\
%%Base                   & Flying lead type \\
% Operating temperature  & down to -196 C \\
%Pressure endurance (Gauge) & above 2 bar at -196~$^\circ$C \\
Max suppy voltage (anode-cathode) & 2000 V \\
Photocathode Q.E. at 420nm  & $\ge 16$\%  \\
%Photocathode Q.E. at 128nm with TPC coating   & $12 \pm 1 $\%    \\
Photocathode Q.E. surface uniformity at 420nm & within $\pm 5 $\% of mean value\\
% Photocathode Q.E. at 420nm with Pt layer & better than 16\%  \\
%Q.E. uniformity at 420 nm & $\pm 10$\% \\
% over the nominal photocathode  surface (up to 10 cm from the center)\\
Number of dynodes     & 10\\
Typical Gain           &  $1\times 10^7$ at 1500 V\\
%Min Gain               &  $0.7\times 10^7$ at 1500 V\\
%%%%Max applied Voltage$^*$  & 1750V \\
Nominal anode pulse rise time$^*$ & $\le 4$~ns\\
%%Nominal pulse trailing edge$^*$ & 10~ns\\\bibliography{bibfile}
Nominal P/V ratio$^*$ & 2.5 \\
Max. dark count rate$^*$   & 5000~s$^{-1}$ \\
%Nominal electron transit time & 54~ns \\
Max. transit time variation   & 2.5~ns (center-border)\\
%Max. transit time variation   & 1.85~ns (center-border)\\
Transit time spread (RMS) & 0.7~ns\\
%SER Peak-to-Valley Ratio min. & 2 at $G =1\times 10^7$ \\
Pulse linearity variation$^{**}$ & $\le 10$\% up to 150 phe\\
                                 & $\le 50$\% up to 300 phe\\
\hline
$^*$ Values for $G =1\times 10^7$ & \\
$^{**}$ Values for $G =1.3\times 10^7$ at 87~K~\cite{Babicz:2019pll} & \\
%\bottomrule 
\end{tabular}
\end{table}

For the upgrade of the ICARUS T600 scintillation light detection system,
400 Hamamatsu R5912-MOD PMTs were procured in 2016.
All the samples were tested at room temperature and 60 of them were also characterized 
at cryogenic temperatures, in liquid argon bath during a test campaign carried out at CERN
%in 2016 and 2017~\cite{Babicz:2018svg,Raselli:2016zzl}. The main results can be summarized as:
in 2016 and 2017~\cite{Babicz:2018svg}. 
%The main results can be summarized as:
%\begin{itemize}
%\item Signal shape characteristics (leading edge $\approx 4$~ns; FWHM $\approx 6$~ns; 
%trailing edge $\approx 10$~ns) suitable for fast triggering application;
%\item Gain variations at cryogenic temperature (down to 10\% of the nominal values) 
%always compensated by an increase of less than 150~V of the power supply voltage;
%\item Dark counts and noise ($\approx 5$~kHz) compatible
%%with the correct operation of this PMT model in LAr bath;
%\item Good uniformity of the response of the cathode all over the sensitive window;
%\item No mechanical or cracking problems observed at cryogenic temperature.
%\end{itemize} 
All the 400 PMTs were rated compliant with the requirements
for installation in the T600.

Each device was equipped with a proper base voltage divider directly welded on the
PMT flying leads. The base circuits, entirely passive, were manufactured
with SMD resistors and capacitors able to withstand the LAr temperature.
%{\color{BlueViolet} 
The reference voltage distribution ratio is the standard
recommend by Hamamatsu. A particular care was devoted to the choice of damping
resistors to improve the PMT time response.  %}
Detailed design and specifications are presented in reference~\cite{Babicz:2018svg}.

\subsection{Wavelength shifter deposition}
\label{evaporation}

The glass window of this PMT model is not transparent to the scintillation light produced by liquid
argon.
The sensitivity to vacuum-ultra-violet (VUV) photons was
achieved by depositing a layer of a proper wavelength shifter on the PMT windows.

1,1,4,4-Tetraphenyl-1,3-butadiene, or TPB, is
an organic fluorescent chemical compound generally
used as wavelength shifter, thanks to its extremely high efficiency to convert ultra-violet photons
into visible light. To obtain effective TPB layers on a large number of PMTs, ensuring at the same time a high
repeatability and reliability of the operation, a dedicated thermal evaporator was instrumented and a
specific evaporation procedure was defined~\cite{Bonesini:2018ubd}. %\cite{Spanu}.

%{\color{BlueViolet} 

\begin{figure}[!t]
\centering
\includegraphics[width=0.5\columnwidth]{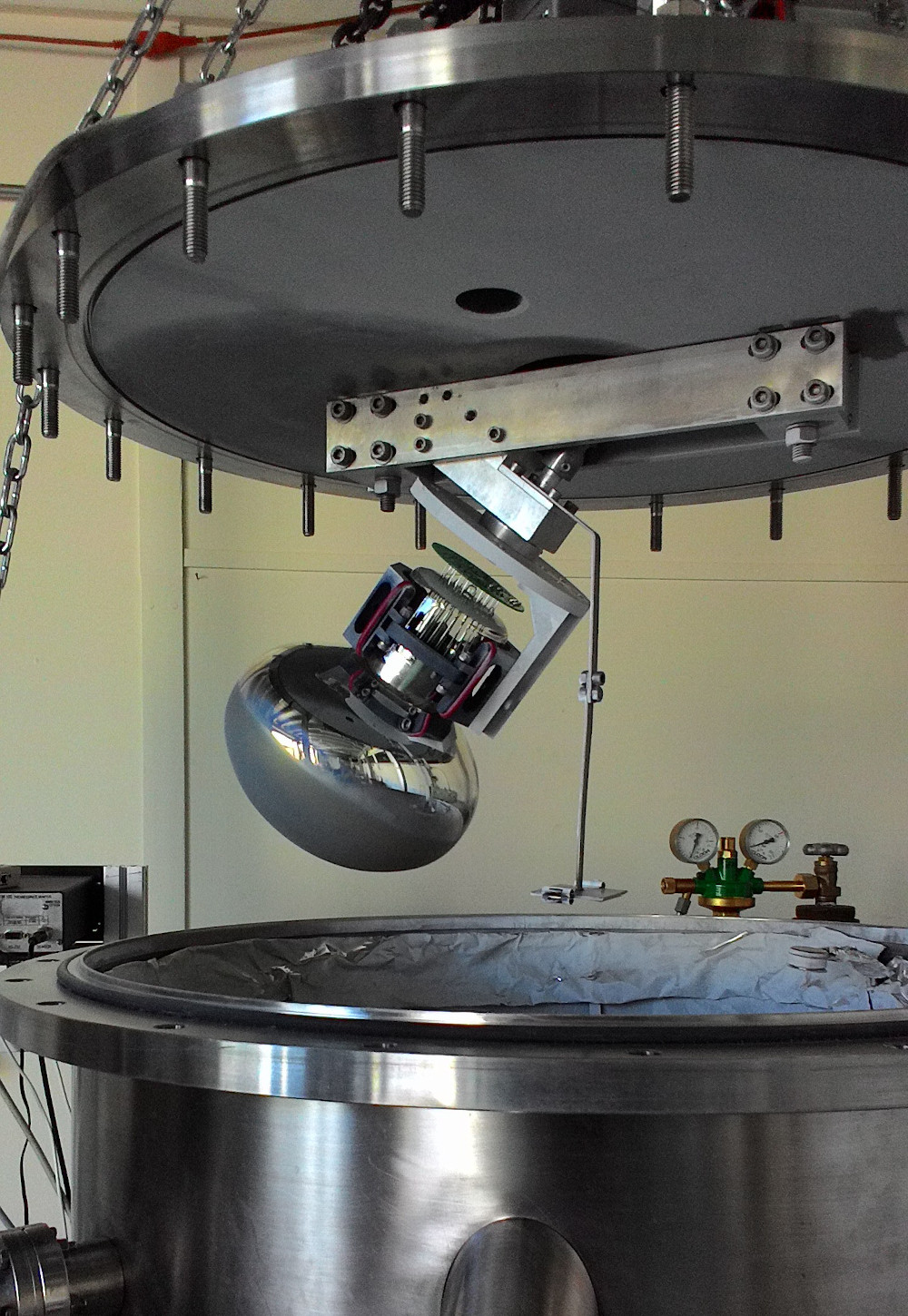}
\caption{Picture of the instrumented evaporator with a PMT fastened on the rotating support.}
\label{fig_evaporator}
\end{figure}

The thermal evaporator consists of a vacuum chamber, 68 cm high and 64 cm diameter, 
closed at both sides by means of two large flanged plates\footnote{The
thermal evaporator and the optical test system cited in this paper were funded by the italian INFN
(“Istituto Nazionale di Fisica Nucleare”) and MIUR (“Ministero dell’Istruzione, dell’Università e
della Ricerca”) within the PRIN (“Progetto di Rilevante Interesse Nazionale”) program.}. 
%thermal evaporator and the optical test system cited in this paper were realized at the
%University of Pavia within the Italian PRIN (“Progetto di Rilevante Interesse Nazionale”) program.}.
The PMT to be coated is fastened to a specific rotating support looking downwards and inclined of
$40^\circ$ angle with respect to the vertical direction, as shown in figure~\ref{fig_evaporator}. 
The rotating structure, fixed below the chamber top plate, %cover,
is connected to an external motor by a ferrofluid-based feedthrough allowing a rotation speeds of 10 turns/min.
The vacuum chamber houses a temperature controlled {\it Knudsen cell}, placed on the bottom plate at a distance
of about 14~cm below the PMT surface.
For each deposition the cell crucible was filled with about 0.8~g of TPB and left to
evaporate at a temperature of $220^\circ$ for about 10~min, yielding a uniform TPB coating 
of about 220~$\mu$g/cm$^2$ on the PMT sensitive surface.
This density value was proven to guarantee a high conversion efficiency and the absence of 
adhesion instabilities on sand-blasted glass at cryogenic temperatures after immersion in LAr~\cite{BENETTI200389}.

\begin{figure}[!t]
\centering
\includegraphics[width=0.65\columnwidth]{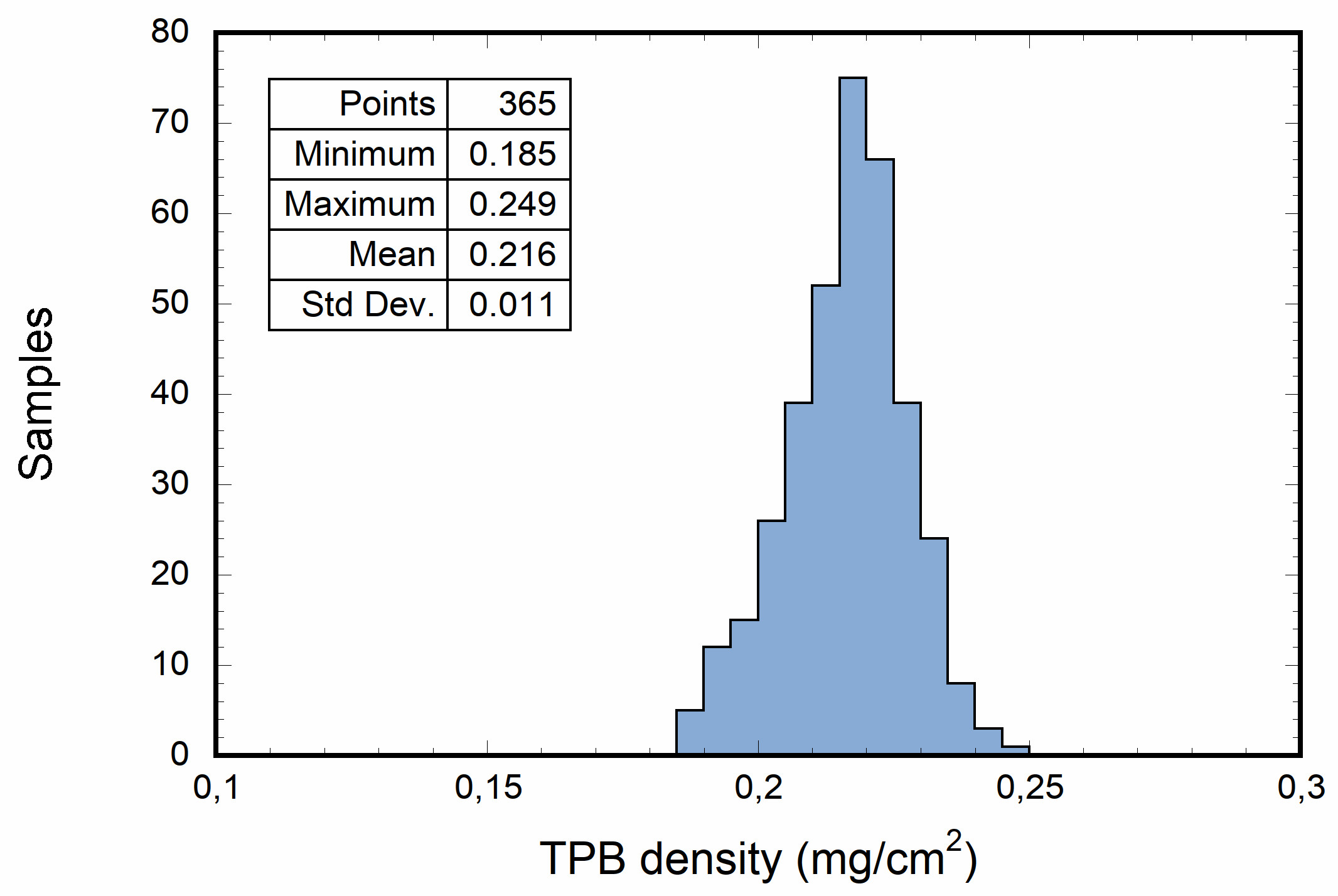}
\caption{Distribution of the resulting TPB coating densities.
Each sample is related to a PMT evaporation run of the series production.
In addition to the needed 360 PMTs, 5 more samples were coated as spare units.}
\label{fig_TPB}
\end{figure}

%Simulation results and experimental tests demonstrate 
The effectiveness of this technique from the point of view of deposition uniformity
and light conversion efficiency was validated by simulations and experimental tests
before being accepted for the TPB coating of the 360 Hamamatsu R5912 of
ICARUS T600 light detection system.
%The adopted technique permitted to perform deposition
%in a short time allowing 
The treatment of a total of 365 PMTs was carried out at CERN Technology Department
in around 120 working days.
The distribution of the resulting TPB coatings is shown in figure~\ref{fig_TPB}.

\begin{figure}[!t]
\centering
\includegraphics[width=0.49\columnwidth]{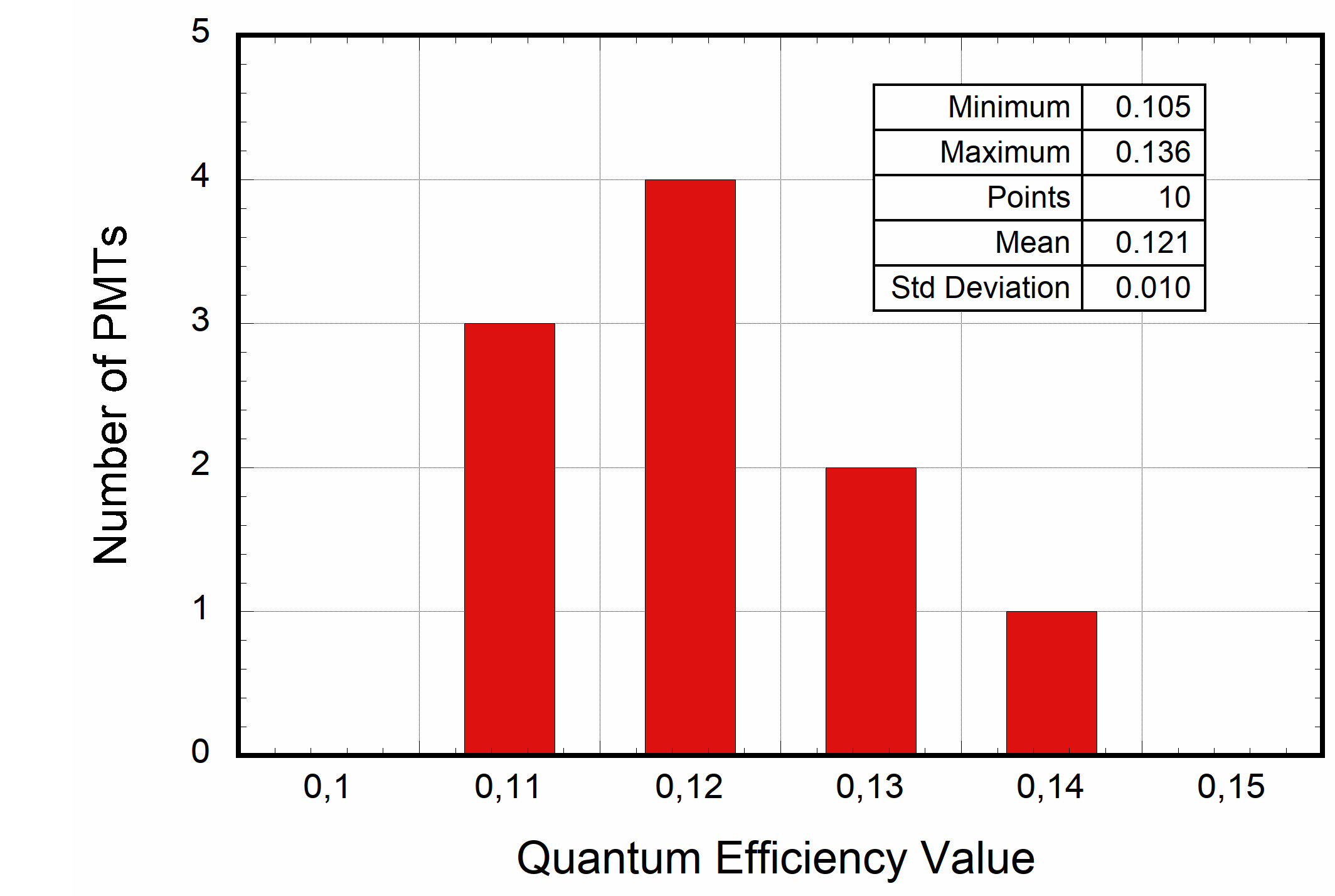}
\includegraphics[width=0.49\columnwidth]{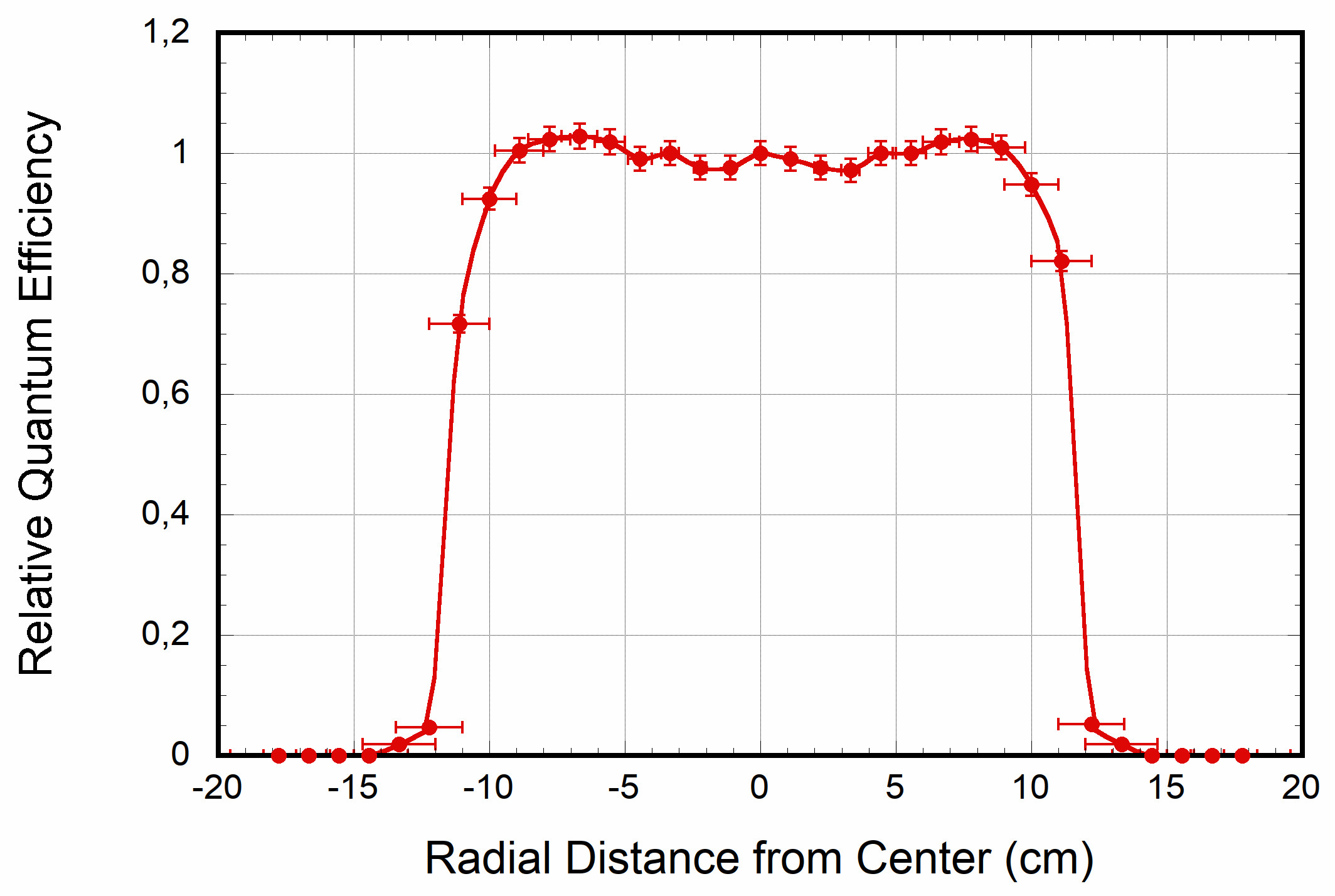}
\caption{(Left) Distribution of quantum efficiency resulting from the
measurement of 10 PMT samples after the TPB coating by evaporation. (Right) Example of measurement of
quantum efficiency variation as a function of the radial
distance from the center.}
\label{fig_QEplots}
\end{figure}

The effective value of the quantum efficiency at $\lambda = 128$~nm and its uniformity as a function
of the position on the photocathode window was measured by means of an optical test system
on 10 PMT samples. 
The quantum efficiency was evaluated by comparing the currents given by the PMT
under VUV illumination and by a reference calibrated photodiode\footnote{
Light at $\lambda = 128$~nm is generated by a 30~W deuterium lamp (McPherson 632) and selected by a
monochromator (McPherson 234/302). The reference is a NIST calibrated photodiode.}. 
Values are
distributed in the $0.11 \div 0.15$ range, with an average value of
0.12, while for each PMT the uniformity results to be within $\pm 5$~\% around the mean value, as shown
in figure~\ref{fig_QEplots}, in agreement with the nominal surface uniformity at 420~nm without TPB.
%}

\subsection{Deployment and installation of the PMTs} %behind the wire planes}

\begin{figure}
\center
\includegraphics[width=0.90\columnwidth]{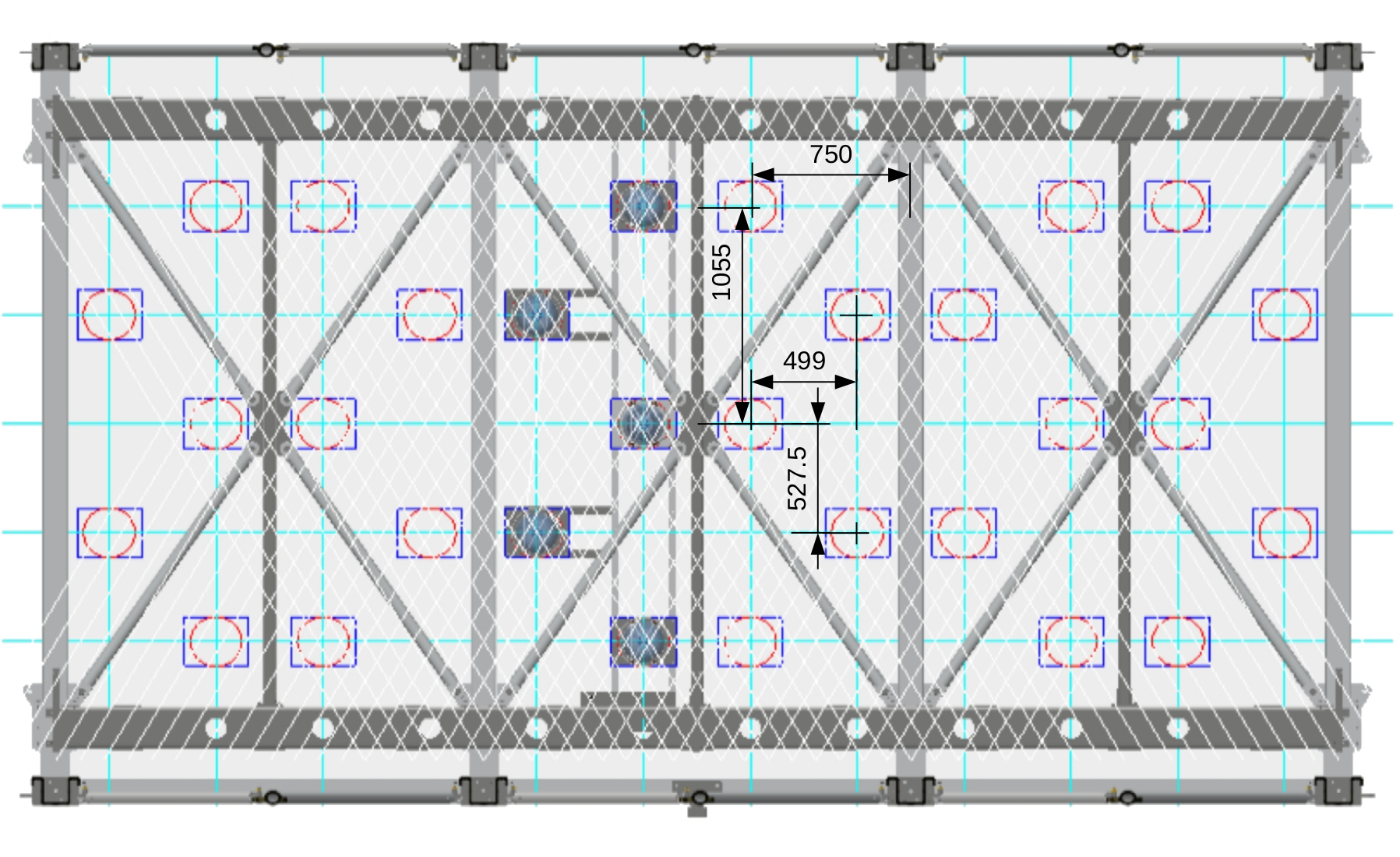} 
\caption{Drawing of the PMT positioning behind the wire planes. Units of measurement are millimiters.
Three  frame sectors are displayed.}
\label{fig_scheme} 
\end{figure}

Each of the two T600 LAr cryostats features a mechanical
structure that sustains the different internal
detector subsystems and the control instrumentation.
The three wire-planes of each TPC are held by a sustaining/tensioning frame
positioned onto the longitudinal side walls of the cryostat.
The stainless-steel supporting structure has dimensions of 19.6~m in
length, 3.6~m in width and 3.9~m in height, subdivided in
9 sectors, 2~m long each.
PMTs are located in the 30~cm space behind the wire planes,
10 unit for each frame sector, as shown in figure~\ref{fig_scheme}.

The PMTs are mounted onto the mechanical
frames by means of PEEK\textsuperscript{TM} holders in form of slabs with dimensions
of $350 \times 250$~mm$^2$, 10~mm thick, held up by stainless-steel bars
3~m long, as shown in~\ref{fig_scheme}. 
The support system allows the
PMT positioning behind the collection planes
$\approx 5$~mm far from the wires. 

\begin{figure}
\center
\includegraphics[width=0.45\columnwidth]{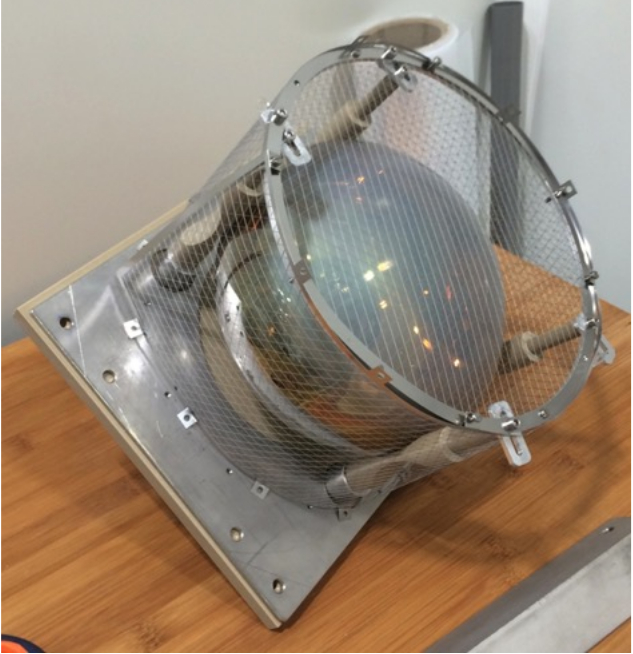} 
\includegraphics[width=0.45\columnwidth]{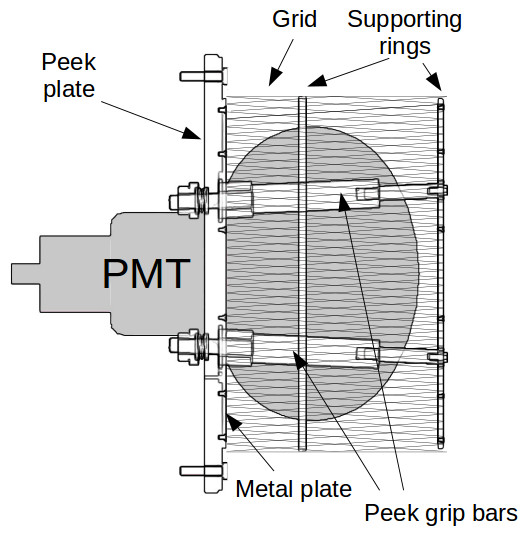}
\caption{Picture (left) and CAD drawing (right) of the PMT support.}
\label{new_support} 
\end{figure}

\begin{figure}
\center
\includegraphics[width=0.45\columnwidth]{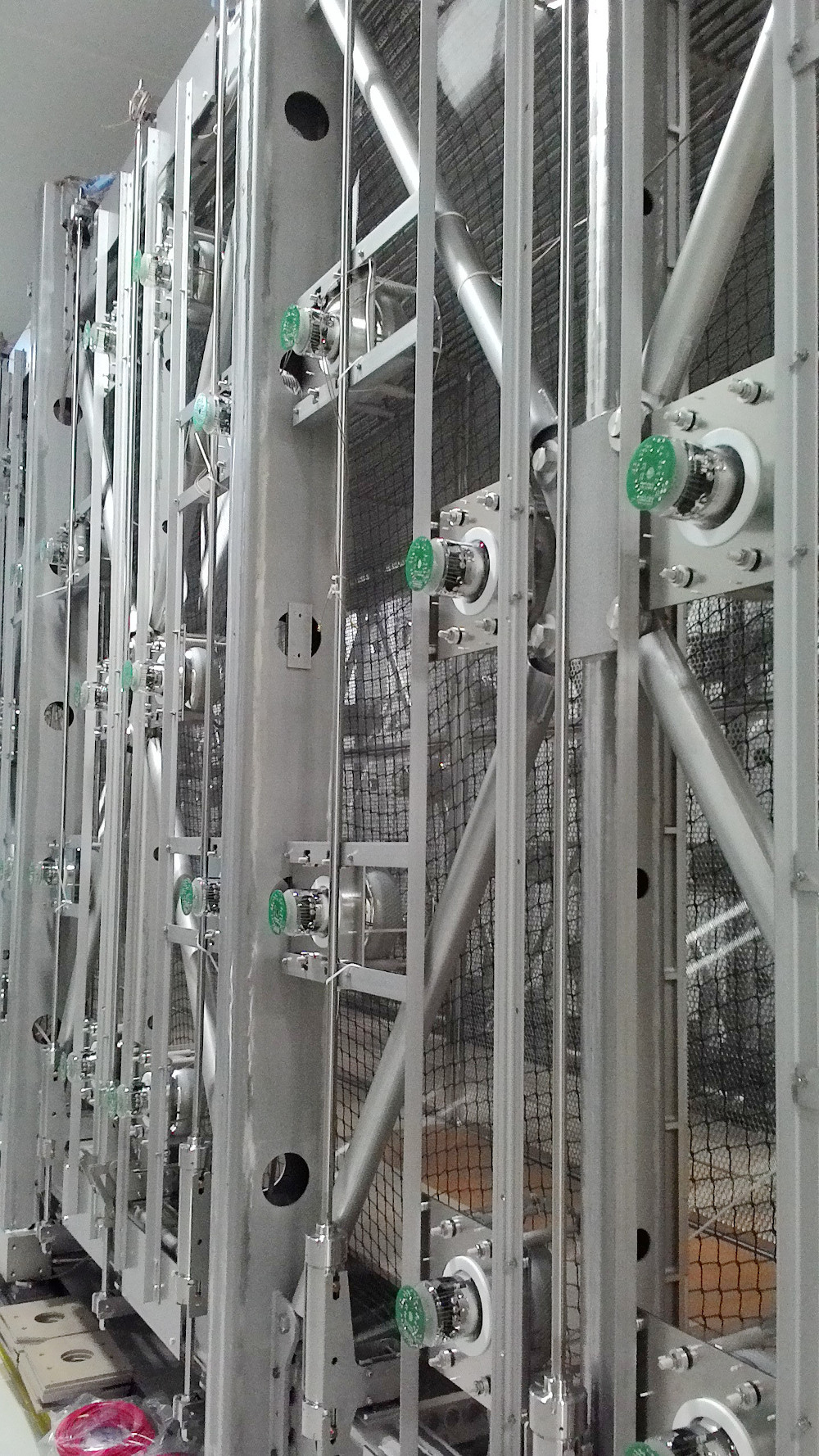} 
\hfill
\includegraphics[width=0.45\columnwidth]{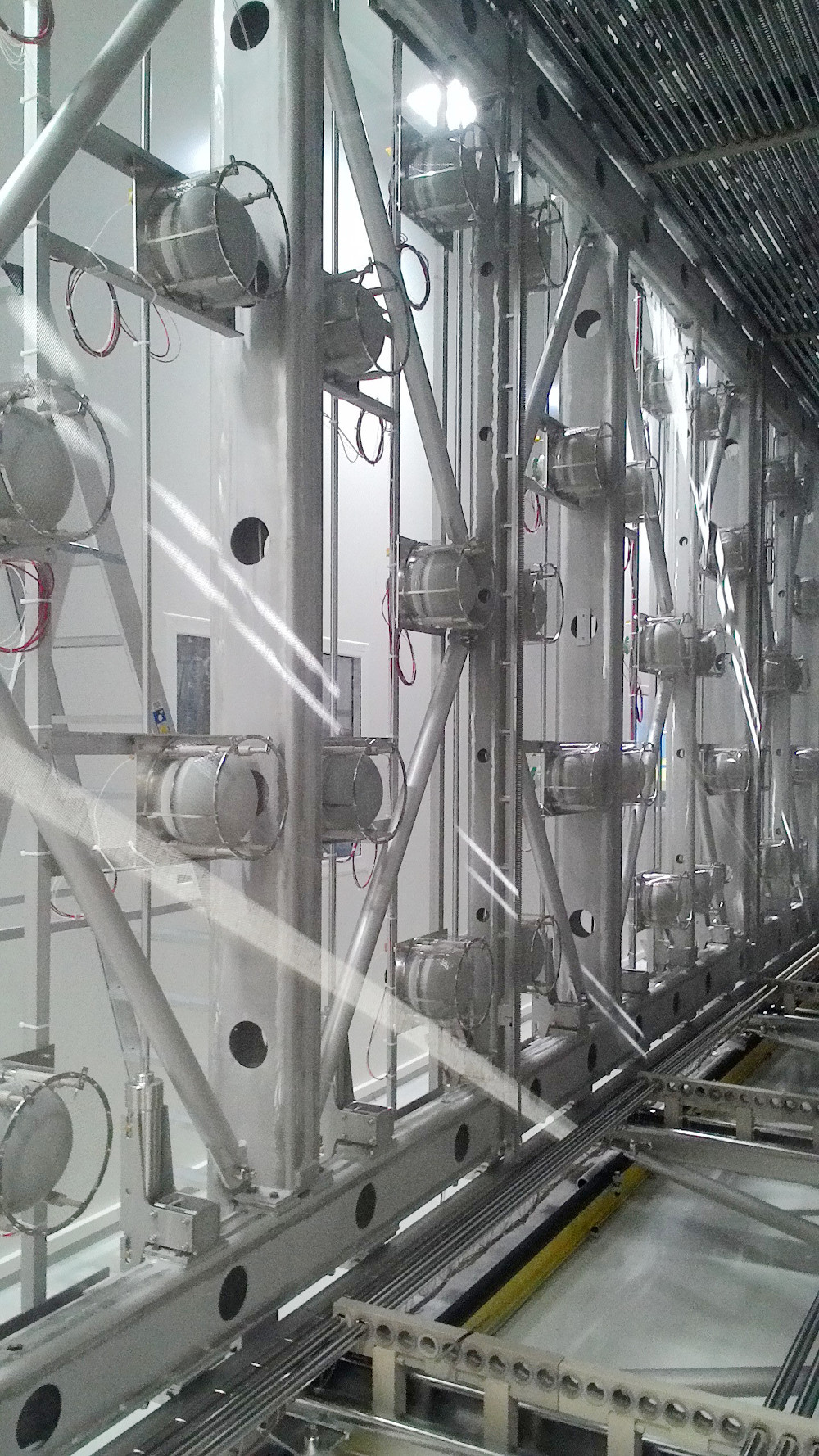}
\caption{Picture of the PMT supporting system.}
\label{holding} 
\end{figure}

The electron multiplication process inside the PMT gain region can induce a fake signal on the wire plane areas facing them. These induced signals appear as fuzzy smeared blobs on a group of wires spanning the PMT diameter.
The effect was witnessed continuously on ICARUS data from LNGS run, but it was negligible and easily identifiable, given the low number of PMTs and tracks per event. At Fermilab, the larger number of PMTs, coupled with higher background track multiplicity, calls for a mitigation of this phenomenon.
A stainless steel grid cage is mounted %between the slab 
around each device as shown in figure~\ref{new_support}. The cage is connected to the detector ground,
as well as the wire polarization system.
Pictures of the supporting system  
are shown in figure~\ref{holding}.

\subsection{PMT cabling and feedthrough installation}

A negative power supply is adopted and two independent coaxial cables are used to provide
each PMT with high voltage and to read out the anode signals. Signal cables are RG316/U, 7~m
long with a BNC connector on one end. High voltage cables are HTC-50-1-1, 7~m long,  with a SHV connector
on one end. The non-terminated ends of the cables are directly soldered on the PMT bases. 

\begin{figure}
\center
\includegraphics[width=1.00\columnwidth]{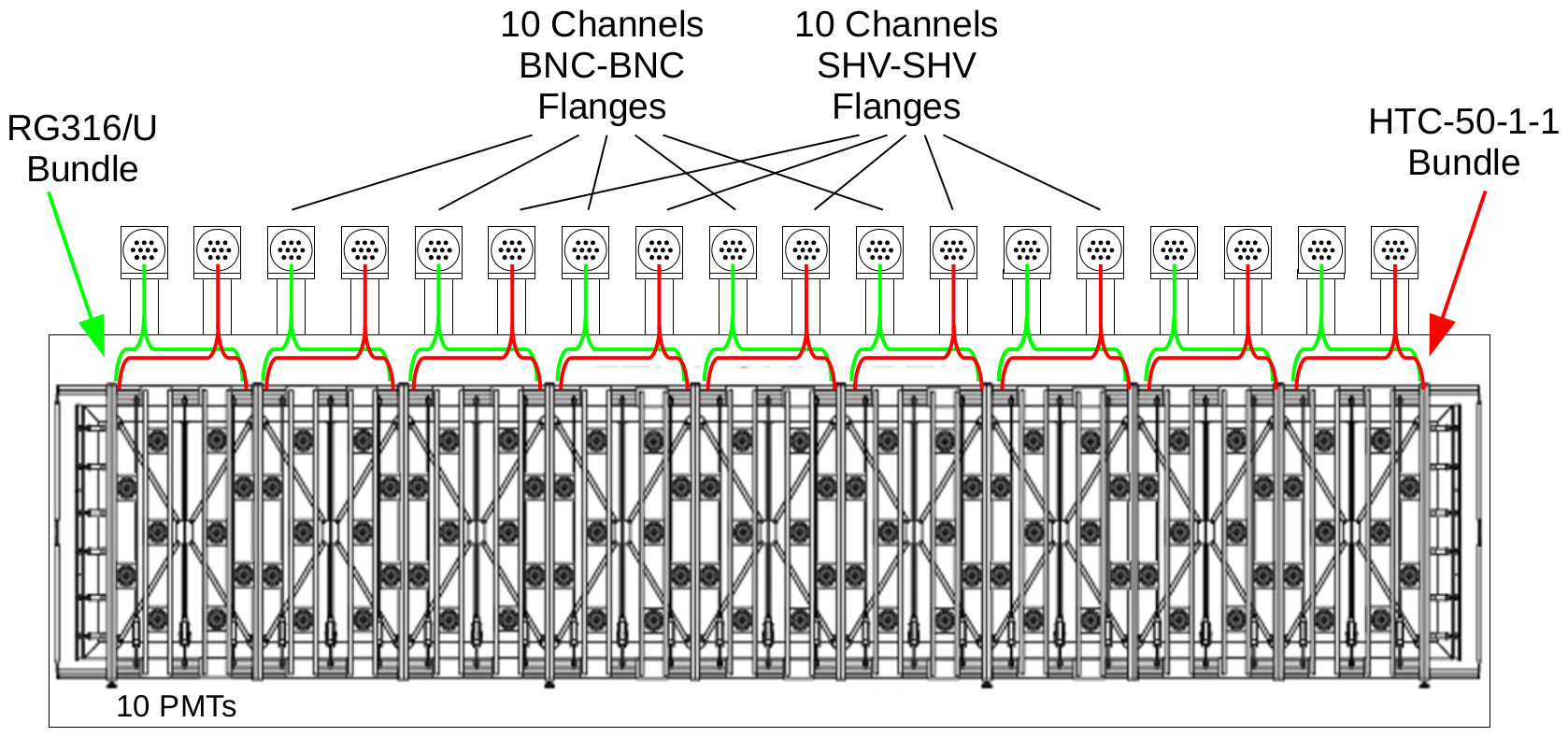} 
\caption{Outline of the PMT cables deployment along the mechanical structure and their distribution on
the top chimneys.}
\label{SchemaHV} 
\end{figure}

\begin{figure}
\center
\includegraphics[width=0.45\columnwidth]{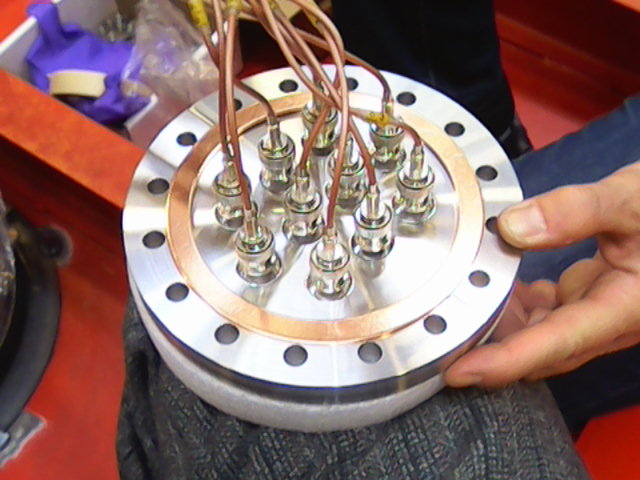} 
\includegraphics[width=0.45\columnwidth]{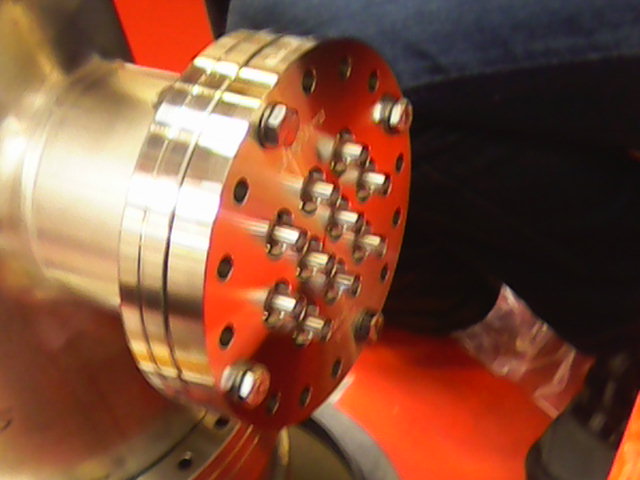} 
\includegraphics[width=0.45\columnwidth]{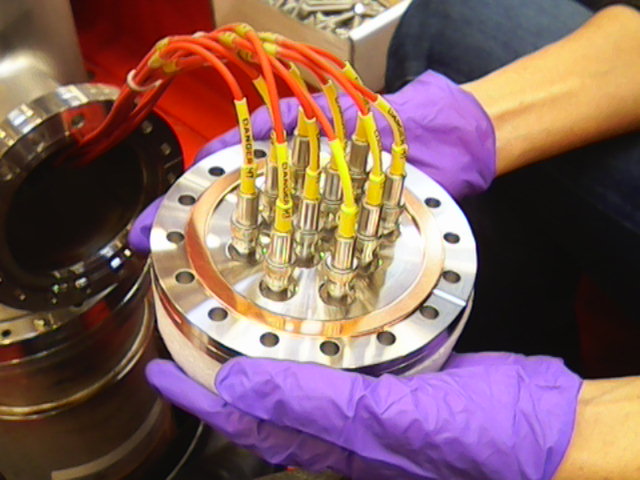} 
\includegraphics[width=0.45\columnwidth]{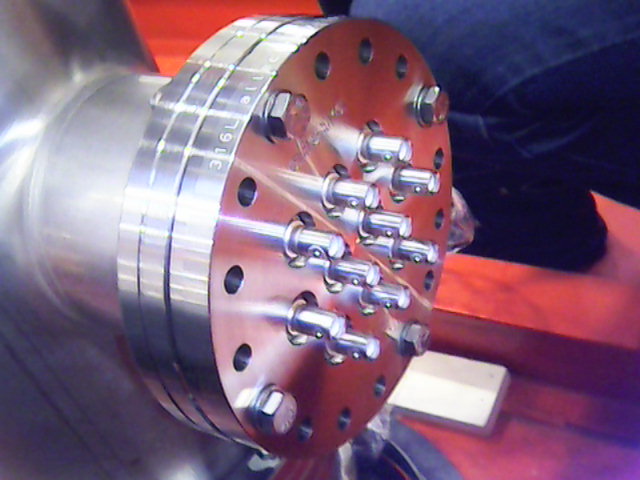}
\caption{Pictures showing the assembly of the PMT flanges. Signal cables and BNC flanges
are shown in the top pictures, while HV cables and SHV flanges are presented in the bottom pictures.}
\label{fig:flanges} 
\end{figure}

For each group of 10 PMTs mounted in the same frame sector, two bundles of 10 signal cables and 10 power
supply cables are deployed along the mechanical structure up to the frame top, as outlined in 
figure~\ref{SchemaHV}.
Each bundle is then driven through a stainless steel chimney (20~cm diameter, 1~m long), vertically
mounted on the detector roof. The top edge of each chimney hosts a set of feedthrough flanges for the interconnection
of various elements of the detector (PMT signal and
power supply, wire signals and biasing, optical fibers and sensors). The PMT flanges are
from Allectra Ltd., each hosting 10 SHV-SHV or 10 BNC-BNC feedthrough connectors mounted
on stainless-steel DN100CF high vacuum flanges. A photo showing the assembly of the PMT
flanges is shown in figure~\ref{fig:flanges}. The main characteristics of the flanges are listed in table~\ref{table2}.

%{\color{BlueViolet} 
The electrical connection between PMT flanges and electronics, located in a building alcove
adjacent to the detector, consists of 360 signal cables (RG316/U with BNC-MCX termination) and
360 high voltage cables (RG58/U with SHV-SHV termination) deployed on cable-trays.
In order to guarantee uniformity among the different channels, all the cables
are 37~m in length. The actual total cable length from PMT base to electronic channel input is
44~m. A detailed study on the effects of the use of these extremely long cables shows a reduction
of the high frequency components of the PMT signal resulting in an increase of the rise-time to 8~ns~\cite{Milind}.
These effects are included in Monte Carlo simulations described in Section~\ref{montecarlo}, to evaluate possible effects
on the system performance.

%The electrical connection between PMT flanges and electronics,
%settled in a  building alcove sideways the detector,
%consists of
%360 signal cables (RG316/U with BNC-MCX termination) 
%and 360 high voltage cables (RG58/U with SHV-SHV termination) deployed on specific cable-traces.
%In order to guarantee uniformity as much as possible
%among the different channels, %for what concerns signal attenuation/distortion and voltage drops,
%all the cables have 37~m length. The actual total cable length from PMT outputs and electronic channel input
%is 44~m.
%%is remarkably long being 44~m. %, in particular for signal propagation.
%% }{\color{Red} 
%A detailed study on the implication of the use of remarkably long cables shows a
%reduction of high frequency components of PMT signal resulting in an increase of rise-time up to 8~ns~\cite{Milind}.
%Results are included in Monte Carlo simulation described in section~\ref{light}, to evaluate 
%possible effects on the system performances.
%%Possible effects on signal shape, such as attenuation/distortion and cable transmission losses,
%%affecting the scintillation light detection performances have been carefully evaluated 
%%%gain changes due to voltage drops,
%%%affecting the scintillation light detection performances,
%%changing on timing parameters and signal amplitude reduction, 
%%have been evaluated, measured 
%%and included in Monte Carlo simulation, as previously described in section~\ref{light}.

%%}

\begin{table}[!t]
\renewcommand{\arraystretch}{1}
% \extrarowheight as needed to properly center the text within the cells
\caption{Flange main characteristics}
\label{table2}
\centering
\begin{tabular}{lll}
\hline
%\toprule
Flange type                    & DN100CF          & DN100CF \\
Number of feedthroughs         & 10               & 10  \\
Connection type (Int/Ext)      & BNC/BNC          & SHV/SHV \\
Impedance                      & 50 Ohm           & 50 Ohm        \\
Shield type                    & Grounded         & Grounded\\
Voltage                        & 1000 V           & 6000 V \\
Min. temperature               & $-200^\circ$C    & $-200^\circ$C\\
Vacuum                         & UHV              & UHV \\ 
\hline
%\bottomrule 
\end{tabular}
\end{table}

%{\color{BlueViolet}
\subsection{PMT electronics}

PMT electronics is designed to allow continuous read-out, digitization and independent
waveform recording of signals 
%from each PMT. %The waveform recording is based on CAEN
%V1730B digitizers. 
%The waveform recording of signals
coming from the 360 PMTs
of the light detection system. This operation is performed by 24 V1730B digitizers. 
% on a 15-channel modularity. A total of 24 digitizer modules are assembled 
%to record the signal coming from the 360 PMTs.
Each module consists of a 16-channel 14-bit 500-MSa/s FLASH ADC.
%{\color{BlueViolet}
The 2 Vpp input dynamic range well fits the PMT response in terms of linearity and saturation
(see table~\ref{table1}). 
In each board 15 channels are used for the acquisition
of PMT signals, while a channel is left for possible future implementations. %, such as the acquisition of
%the trigger signal. 
During the acquisition, data stream of each channel is continuously written every 2~ns in a circular memory
%During the acquisition, every 2~ns the amplitude of each channel is sampled data stream of each channel is continuously written every 2~ns in a circular memory
buffer of 5kSa, corresponding to 10~$\mu$s\footnote{The total memory available for each channel is 5.12 MSa, divisible into a maximum of 1024 buffers.},
allowing %for each event, 
the recording of both components of the LAr
scintillation light, i.e. photons from fast %(time of the order of nanoseconds) 
and slow %(time of about 1.6$\mu$s) 
decays of exited excimers to ground state, as described in Section~\ref{light}.
%When a common acquisition trigger pulse
%(common to all the channels for each board) 
When a boards receives an external trigger request, 
the active buffers are frozen, writing operations are moved to the next 
available buffers and stored data are available for download via optical 
links\footnote{Data read out is based on the CAEN proprietary CONET2 (Chain162
able Optical NETwork) protocol allowing up to  80~MB/s data transfer.}.
%the DAQ. %%This configuration guarantees no
%dead time, until the maximum DAQ throughput is reached. A common acquisition trigger signal
% (common to all the channels) is fed externally via the front panel TRG-IN input connector causing
%134 all the board channels to acquire an event simultaneously.
The amplitude of the prompt signals from the fast component of the scintillation light, without any
pulse integration, is exploited for trigger purposes.
%
%
%Trigger pulses are generated by the ICARUS Trigger System
%every time a ionizing interaction is recognized in the detector
%on the base of information coming from neutrino beams and other apparatus subsystems. %~\cite{TRIGGER}.
%which will be 
%described in dedicated papers.
%To this aim  trigger-request patterns are provided by the V1730NB digitizers when input signals goes under/over 
%programmed thresholds.
%
%when input signals goes under/over 
%programmable thresholds. These patterns are logically processed by the ICARUS Trigger System to identify
%using information coming from neutrino
%%%%beams and various apparatus subsystems~\cite{TRIGGER}.
%both the scintillation light and beam gate informationr~\cite{TRIGGER}.
%Data read-out is carried out by means of 80~MB/s optical links\footnote{Data transfer is based on the CAEN proprietary CONET2 (Chain162
%able Optical NETwork) protocol.}, one per board.
%and Daisy-chain capability.
%Although it whould be possible to connect up to 8 modules to a single optical-link channel, in
%order to exploit the available data acquisition bandwidth and ensure the maximum data transfer rate
%during read-out, an independent optical-link channel is assigned to each module, according to the
%scheme shown in figure ??].
To this aim,
V1730B boards generate a pattern of digital pulses
(200~ns, LVDS logic standard)    %) \footnote{LVDS (Low Voltage Differential Signal) logic standard is adopted.} 
mapping the PMT signals exceeding
digitally programmed thresholds,
set to few photoelectrons~\cite{CAEN}.
The ICARUS Trigger System, which will be 
described in dedicated papers, generates a first level
PMT trigger pulse whenever the coincidence of the discriminated 
PMT signals satisfy a defined multiplicity inside 
a neutrino beam gate window.
%, generated in correspondence to the
%expected arrival time of neutrinos in the T600 from an ‘‘early warning’’
%information on the proton spill extraction.

\begin{figure}
\center
\includegraphics[width=0.65\columnwidth]{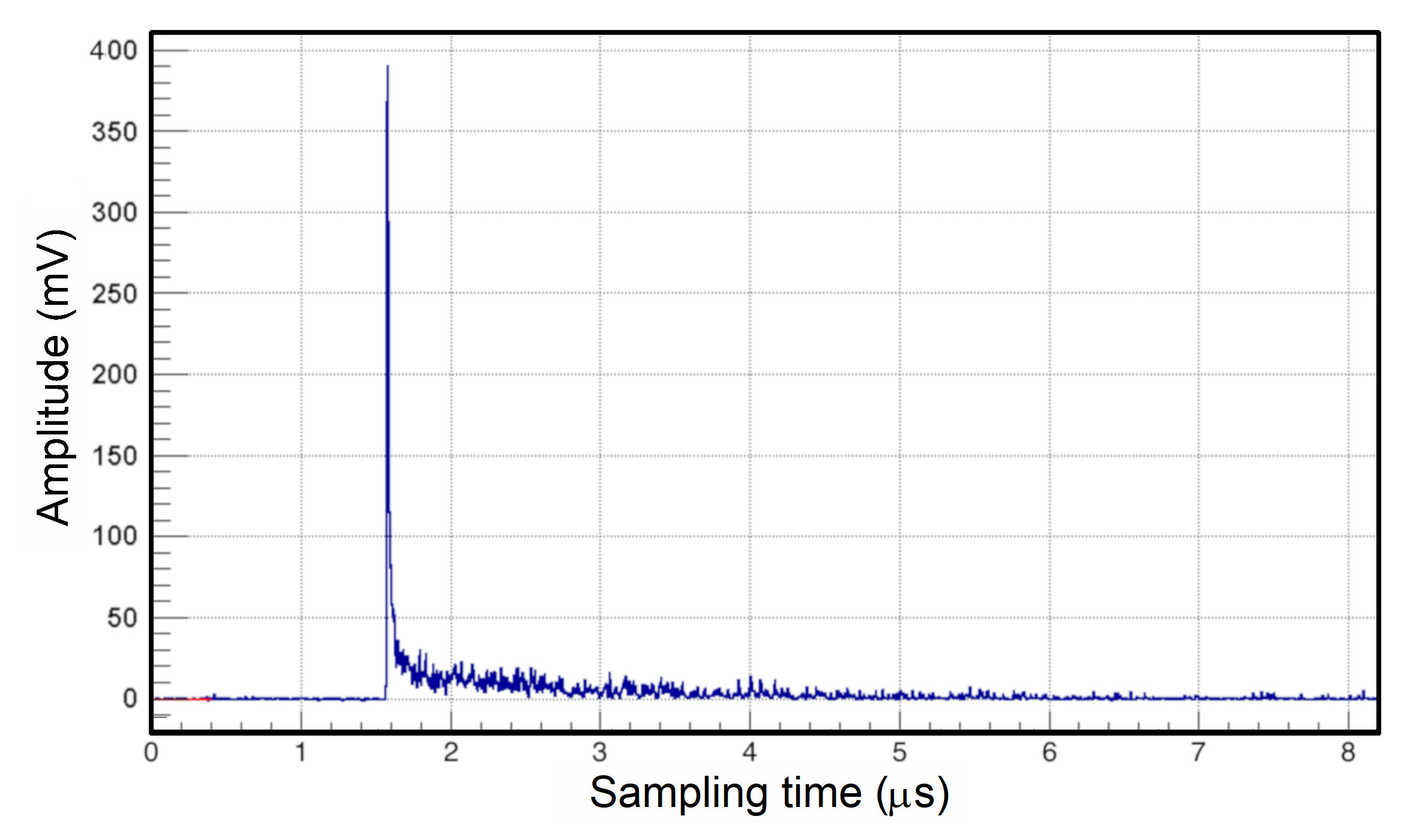} 
\includegraphics[width=0.65\columnwidth]{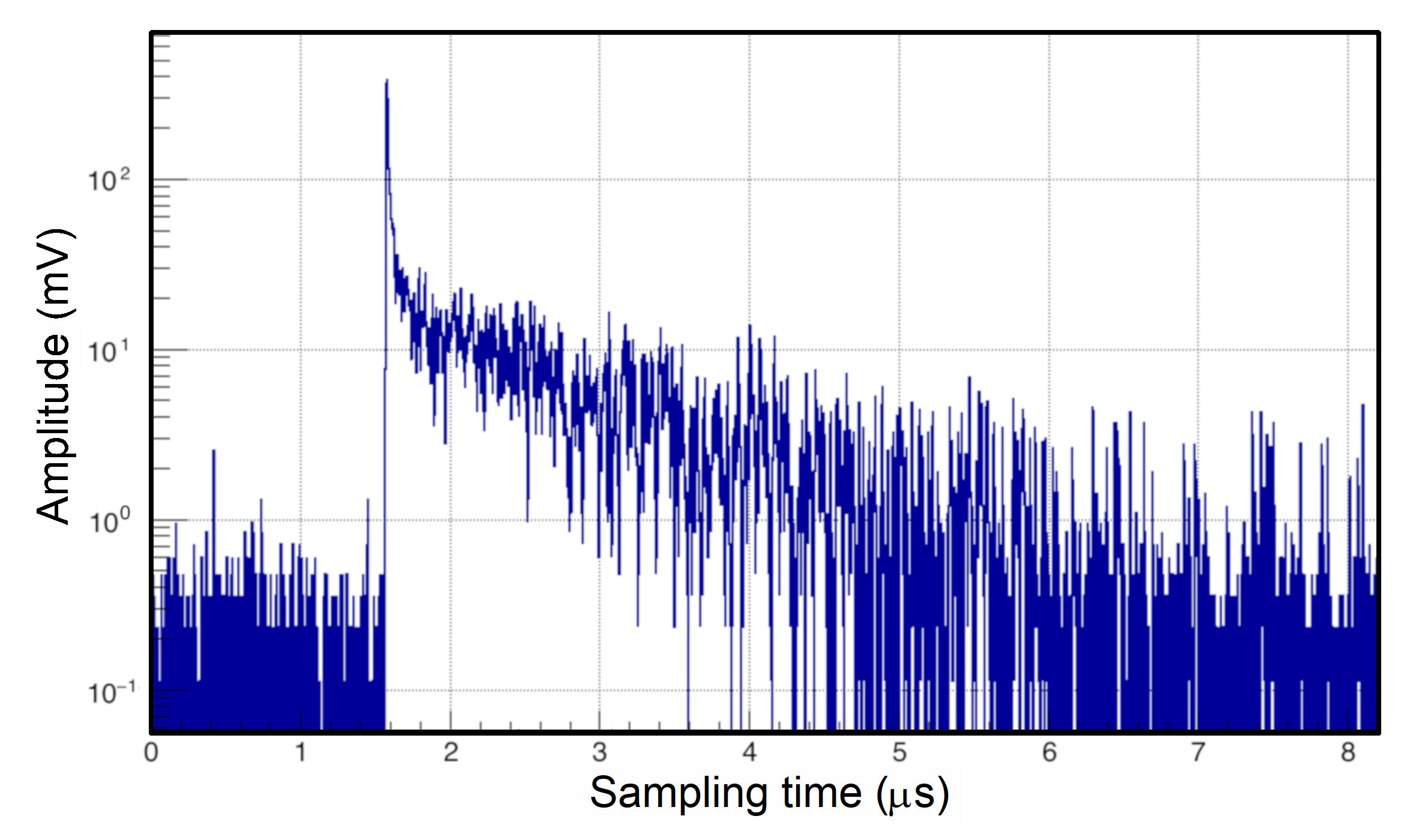} 
\caption{Example of PMT signal (absolute value) recorded with the CERN 10-PMT facility in linear 
({\it Up}\/) and logarithmic  ({\it Down}\/) scale. 
The presence of both the slow and the fast components
of the scintillation light can be noticed.}
\label{fig_single} 
\end{figure}

For generation and distribution of high voltages, the same power supply system designed
by ICARUS for the LNGS run is adopted. For each cryostat, housing 180 PMTs, a primary -2000~V is generated by a  BERTAN 210-02R.
The primary voltage is finely regulated and distributed to 180 PMTs by four CAEN A1932AN boards, 48 channels each, housed in a
CAEN SY1527 crate. The linear technology employed in this system results in extremely low output ripple,
as demonstrated during the ICARUS data taking at LNGS. 
This is a fundamental feature required for the light
detection system to prevent the induction of PMT noise onto the wire planes.
In order to perform a study on the performance of the ICARUS PMT electronics and other detector
subsystems before the final detector operation at Fermilab,
a LAr test facility was instrumented at CERN\footnote{This work was carried out in the framework of the CERN Neutrino Platform WA104/NP01 activities.}~\cite{BABICZ2019162421}.
The apparatus
consists of a 1.5~m$^3$ cylindrical cryostat filled with LAr and instrumented with 10 Hamamatsu R5912-MOD PMTs, 6 of them coated with TPB.
Scintillation light data were taken by exposing the system to cosmic rays. % or using an $^{241}$Am alpha source.
Figure~\ref{fig_single} shows an example of PMT signal taken with this facility, demonstrating the capability of
detecting both components of scintillation light and demonstrating the required  performance of the electronics.

%}

\subsection{Layout of the laser calibration system}

To identify interactions associated to the neutrino beam and reject 
the expected huge cosmic background, the occurrence time of each event should be 
reconstructed with a resolution better than 1~ns. This can be achieved by means of 
a precise determination of the time delay of the response of each PMT,
that may drift 
in time for temperature excursions, power supply variations or other reasons.
The monitoring of the timing values during data-taking
can be accomplished with cosmic rays or by delivering a
fast calibration pulse to each individual channel. 

\begin{figure}
\center
\includegraphics[width=1.0\columnwidth]{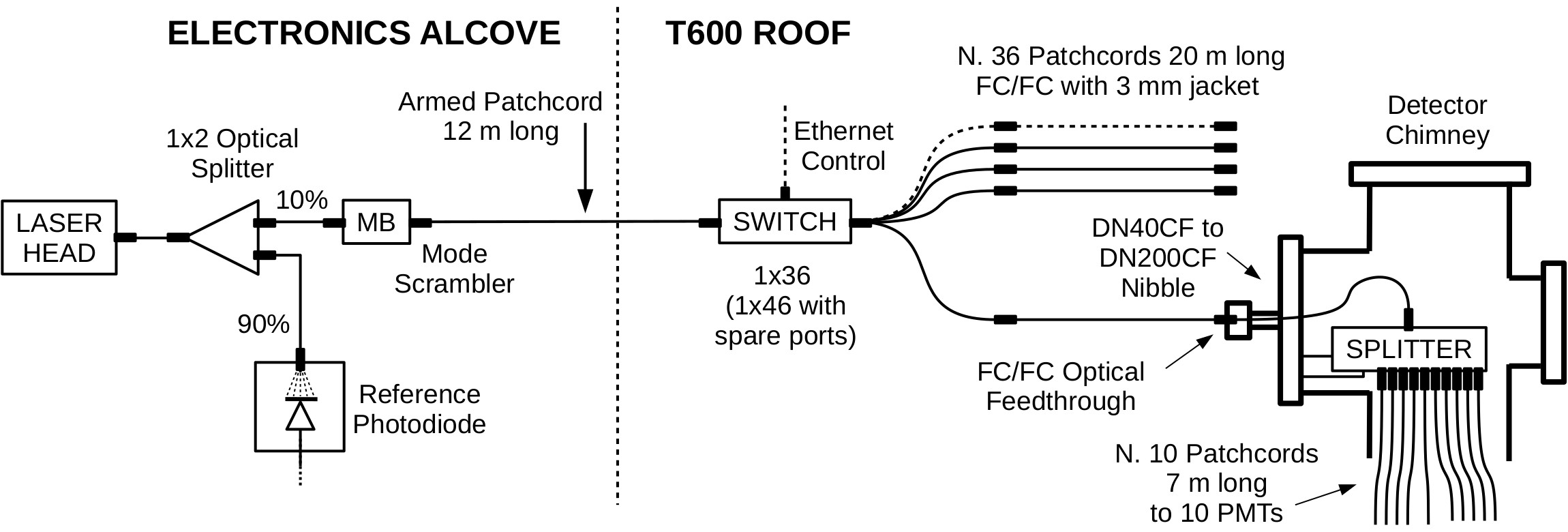} 
\caption{Diagram of the laser calibration system.}
% 1) 405~nm Hamamatsu PLP-10
%laser; 2) $1\times 42$ Agiltron optical switch;
%3) 20~m  long FC/FC patchcords  [36 pcs]; 4) FC/FC UHV feedtrough [36 pcs];
%5) $1\times 10$ fused fiber splitters [36 pcs]; 6) 7~m FC/ferrule injection patch fiber [360 pcs].}
\label{laser} 
\end{figure}

\begin{figure}
\center
\includegraphics[width=0.45\columnwidth]{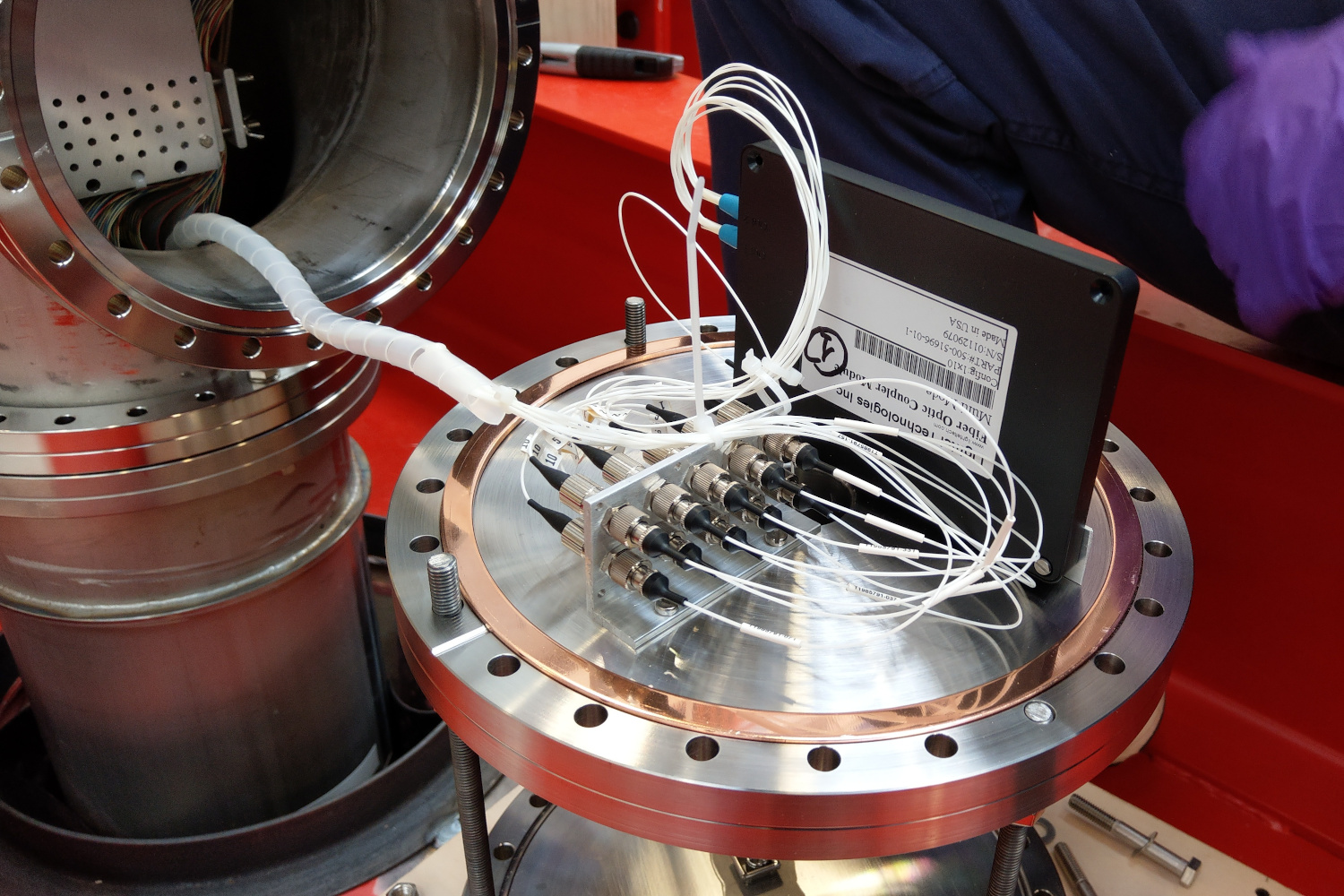} 
\includegraphics[width=0.45\columnwidth]{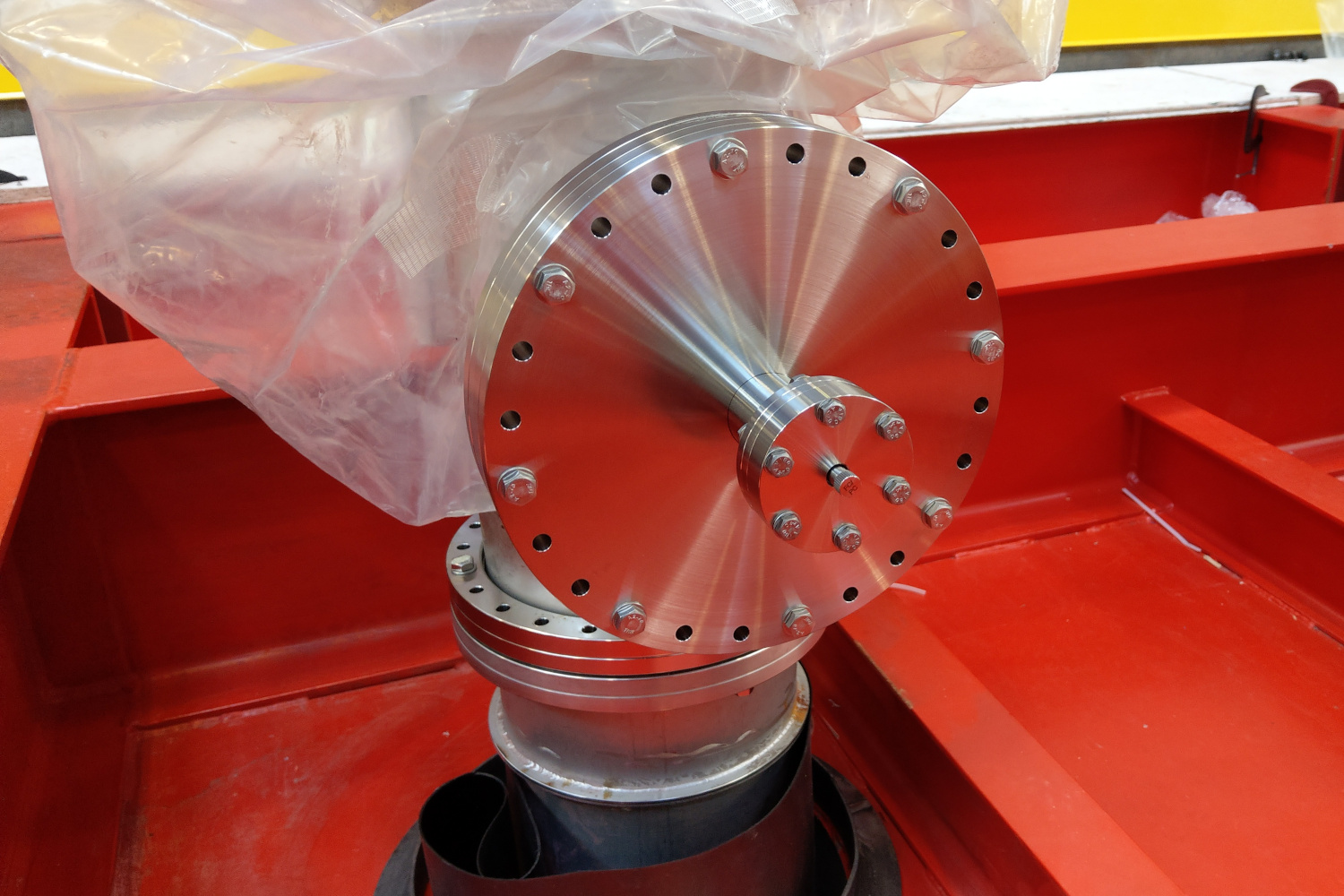} 
\caption{Left picture: DN200CF side of one nibble with a $1 \times 10$ splitter (on the right) and a
10-channel patch panel to connect the internal patches to the outputs of the splitter (on the left).
Right picture: front view of the DN200CF to DN40CF nibble with a mounted FC/FC optical
feedthrough.}
\label{FlangiaLaser} 
\end{figure}

A fast-laser based calibration system has been developed for the time calibration and monitoring
of each PMT channel. 
Its layout is outlined in figure~\ref{laser}, while additional information on the employed components
can be found in reference~\cite{bonesini19}.
Fast light pulses (60~ps FWHM, 120~mW peak power, emission at 405~nm) are generated by a
laser diode (Hamamatsu PLP10) settled in the building electronics alcove.
%sideways the detector. 
%Light pulses are splitted toward (90~\%) a reference photodiode (Thornlabs DET02AFC)
%to monitor the laser stability and toward (10~\%) a ModCon mode scrambler (Arden Photonics) to ensure the same distribution
%of launched modes into the fibers. A 12~m long $50 \; \mu$m armed patch cable from OZ/Optics Ltd. fed a $1\times 46$ optical 
%switch (Agiltron Inc.) placed on the T600 detector roof. An external Ethernet control allows the 
%the selection of the switch channel, 36 of which used to fed
%36 MM (``Multi Mode'') $50 \; \mu$m, 20~m long, optical patch 
%cables from OZ/Optics Ltd., with FC (``Ferrule Connector'') fiber mates. 
%To adapt these flanges to the existing T600
%chimneys, DN40CF to DN200CF adapters are used.
Light pulses feed, through  $50 \; \mu$m patch cables and an optical switch (Agiltron Inc.), 36 optical flanges mounted in the same 36 chimneys 
used for the PMT signal cables, on the opposite site of the BNC flanges.
%These arrive to 36 optical FC/FC
%UHV optical feedthroughs (VACOM  GmbH, $50\; \mu$m core through fibers) on DN40CF flanges.
The main characteristics of the optical flanges are shown in table \ref{table3}. 
\begin{table}[!t]
\renewcommand{\arraystretch}{1}
\caption{Optical flange main characteristics}
\label{table3}
\centering
\begin{tabular}{ll}
\hline
%\toprule
Manufacturer                   & VACOM  GmbH \\
Flange type                    & DN40CF      \\
Connection type (Int/Ext)      & FC/FC \\
inside fiber                   & MM 50 $\mu$m core; NA 0.2 \\
Min. temperature               & $-25^\circ$C    \\
Max. temperature               & $ 75^\circ$C \\
Vacuum                         & UHV              \\ 
\hline
%\bottomrule 
\end{tabular}
\end{table}
%To adapt these flanges to the existing T600
%chimneys, DN40CF to DN200CF adapters are used.
%The optical flanges are
%mounted in the same 36  chimneys used for the PMT signal cables, on the opposite site of the
%BNC flanges.
Inside each chimney, % 36 fused fibers
a $1\times 10$ optical splitter 
(Lightel Technologies Inc.) %made of Corning 50/125 MM fibers)
%attached to the inner side of each CF40 to CF200 adapter
delivers the input laser signal to 10 % a total of 360
($50~\mu$m, 7~m long) injection fibers %\footnote{OZ/Optics MM IRVIS fibers 50/125 $\mu$m with a 0.9 mm hytrel jacket},
deployed along the mechanical
frames, to convey the calibration signal to each
PMT, as shown in figure~\ref{FlangiaLaser}.

\begin{figure}
\center
\includegraphics[width=0.50\columnwidth]{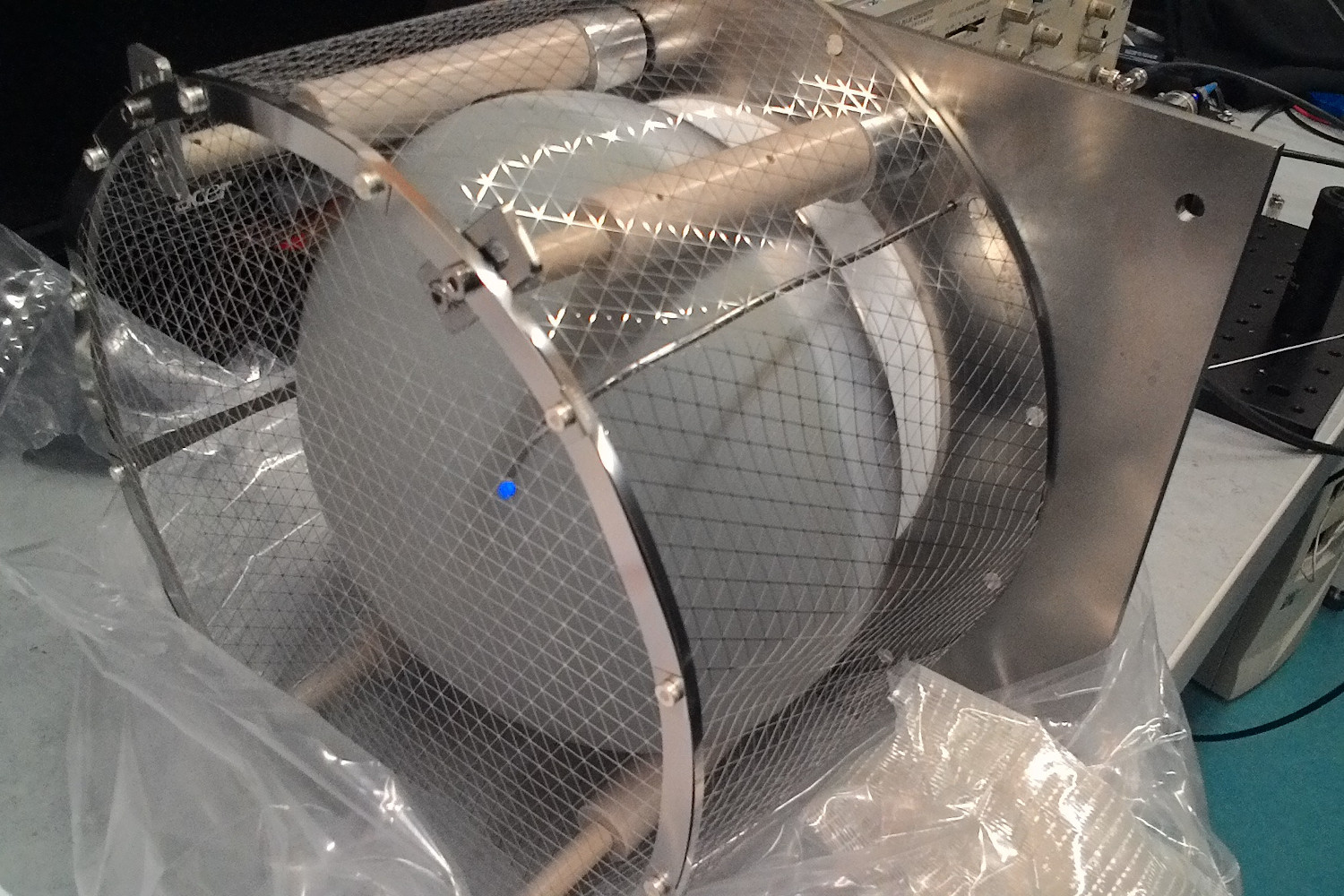} 
\caption{Picture showing the stainless steel pipe which drives the end section of the calibration optical fiber toward
the PMT window. A blue-light laser spot on the PMT surface can be noticed.}
\label{LaserFocus} 
\end{figure}

A specially shaped stainless steel pipe (2.5~mm diameter, 20~cm long), fixed inside the PMT sustaining
structure, drives the end section of each optical fiber and
allows the light focusing on the device windows, as shown in figure~\ref{LaserFocus}.

\begin{figure}
\center
\includegraphics[width=0.49\columnwidth]{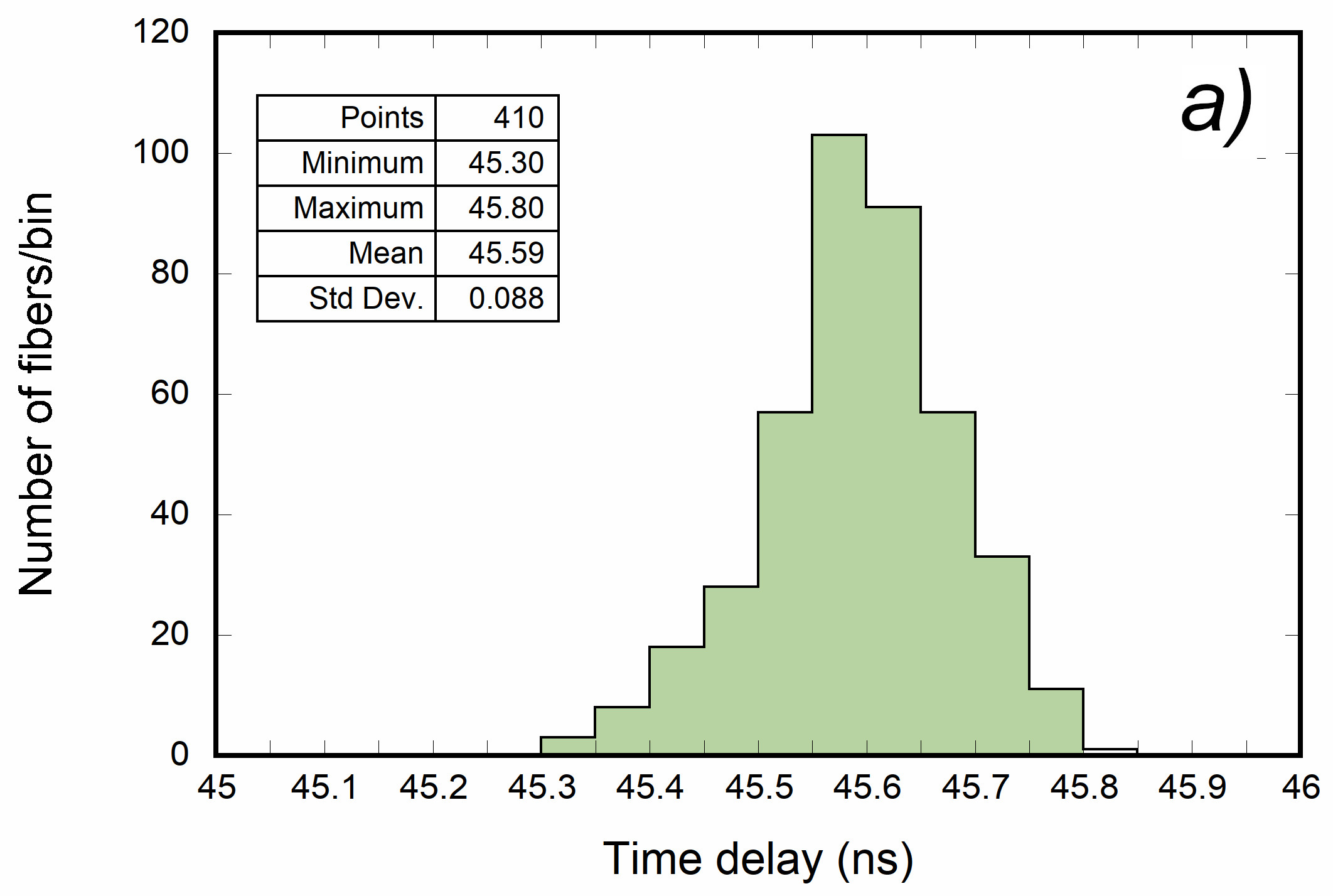} 
\includegraphics[width=0.49\columnwidth]{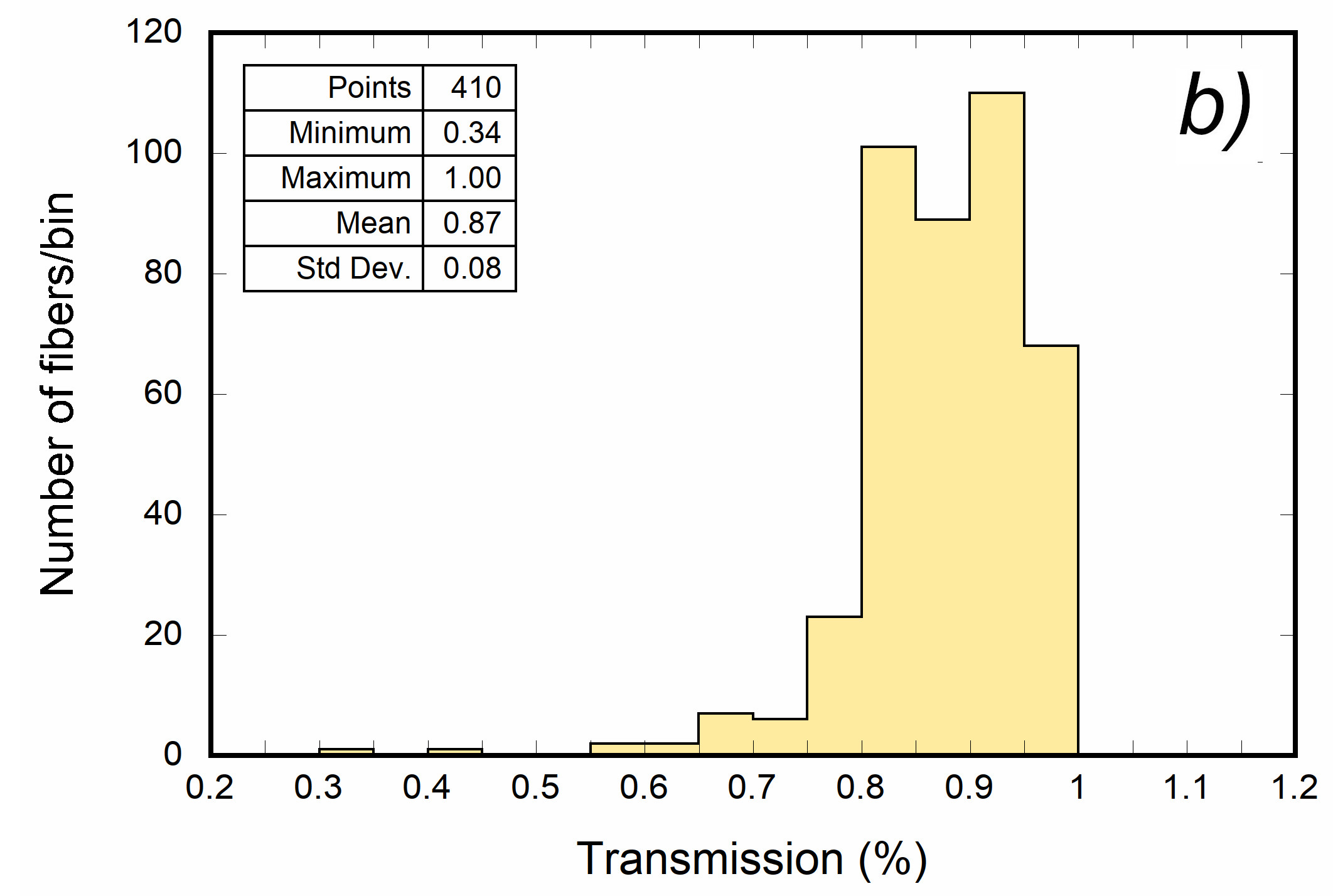} 
\caption{{\it a)}\/ distribution of 7~m injection patches time delay;
{\it b)}\/ distribution of transmission for the same sample.}
\label{mio} 
\end{figure}

The laser pulses should be delivered
to the PMT photocathodes with minimal attenuation and
without deterioration of the original timing characteristics.
Extensive tests were performed on the different components at both room and
cryogenic temperatures,
%, as reported in~\cite{bertoni16,laser_test}, 
to ensure that selected 
items comply with  these requirements~\cite{bonesini19}.
Figure~\ref{mio} shows the distributions of time delay and fraction of transmitted light,
obtained from measurements at room temperature on the initial sample
of 410 injection patches, using the laboratory setup of reference \cite{bertoni16}. 
The delay dispersion over the 7~m cable length is within 90~ps, with an average value of
about 45.59~ns, while the dispersion on the transmission is around $8 \%$ 
with an average value of $\approx 88 \%$. Measurement at cryogenic temperatures (with a LN$_2$ bath) shows that 
transmission and delay are similar to the ones measured at room temperature.

%%%\begin{figure}
%%%\center
%%%\includegraphics[width=0.50\columnwidth]{mio.png} 
%%%\caption{Top panel: distribution of 7m injection patches time delay;
%%%bottom panel: distribution of transmission for the same sample.}
%%%\label{mio} 
%%%\end{figure}

%{\color{BlueViolet}
The final sample of 360 internal injection 
patches was then selected, requiring fibers with the highest transmission and
similar delays. 

The numbers here presented for the laser system only concern the components within the cryostats, 
which are relevant for the purpose of this paper, while the main characteristics of 
the external system are presented elsewhere~\cite{bonesini19}.
%The transmission of the external system, up to the optical
%flange input, was measured after installation at Fermilab as $35.4 \pm 1.6$~\%.
The installation of the various components %(internal system at CERN and external part at Fermilab) 
is still in progress and did not allow a precise calibration  of the entire system on a channel-by-channel basis.
Anyway, the different test results
indicate that the expected 
performances of the laser calibration system, such as an intrinsic time calibration resolution of about 100~ps
on the single PMT channel~\cite{bertoni16}, well fit
the calibration requests for
PMT time equalization  at the level of 1~ns~\cite{Antonello:2015lea}.
The complete description of the laser system performances will be the subject of a forthcoming paper.

%}

\section{Final system tests}
\label{tests}

%The whole light detection system was tested at Fermilab after the installation
%of the detector flanges, once the dark condition inside the cryostats was guaranteed.   
%The aim was to check all the PMT channels, evaluate the PMT performance before the
%cooling down, gain knowledge on how to perform calibration at commissioning stage
%and evaluate the effectiveness of the internal optical-fiber
%light-distribution system.

%{\color{BlueViolet} 

\begin{figure}
\center
\includegraphics[width=0.49\columnwidth]{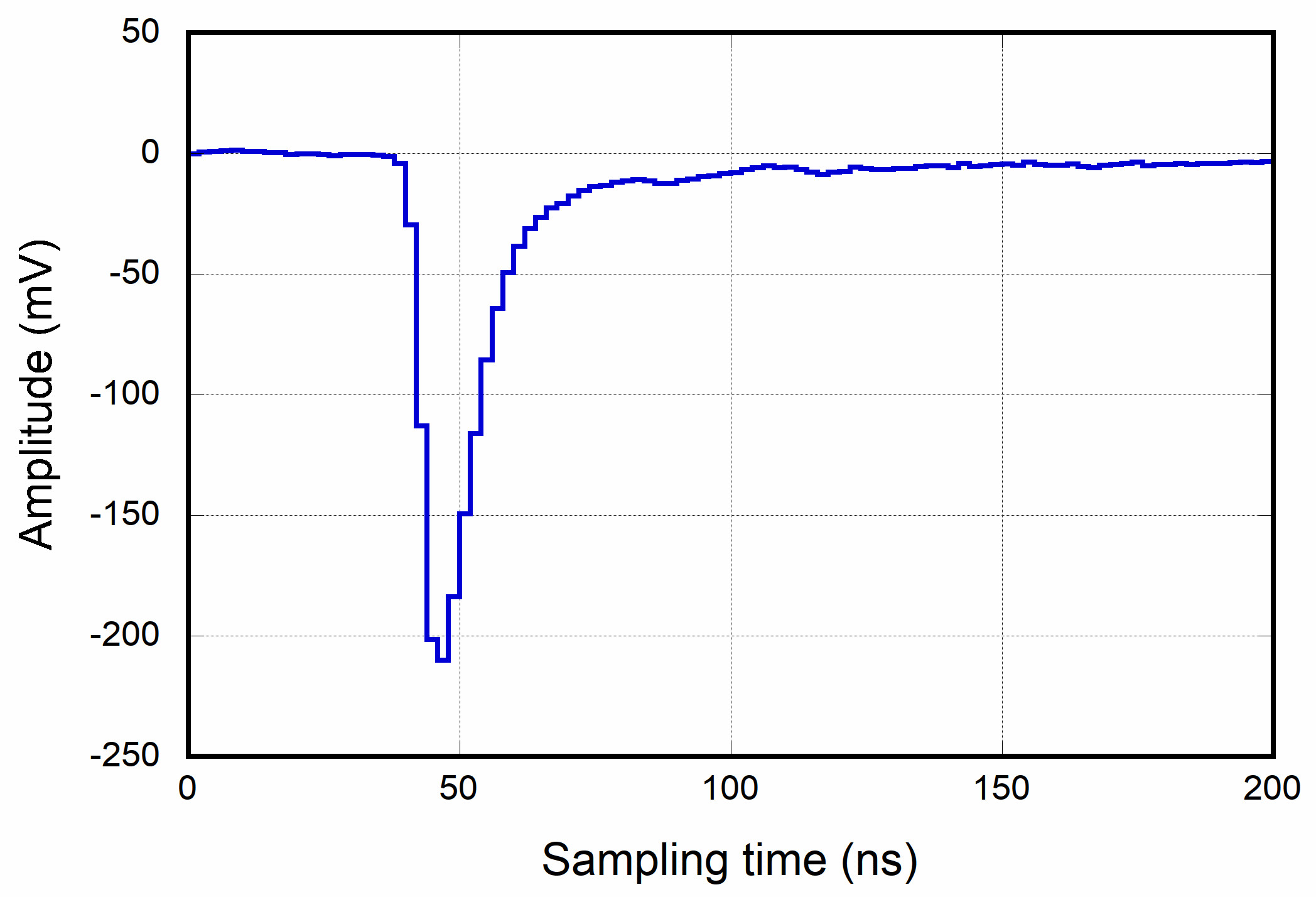} 
\includegraphics[width=0.49\columnwidth]{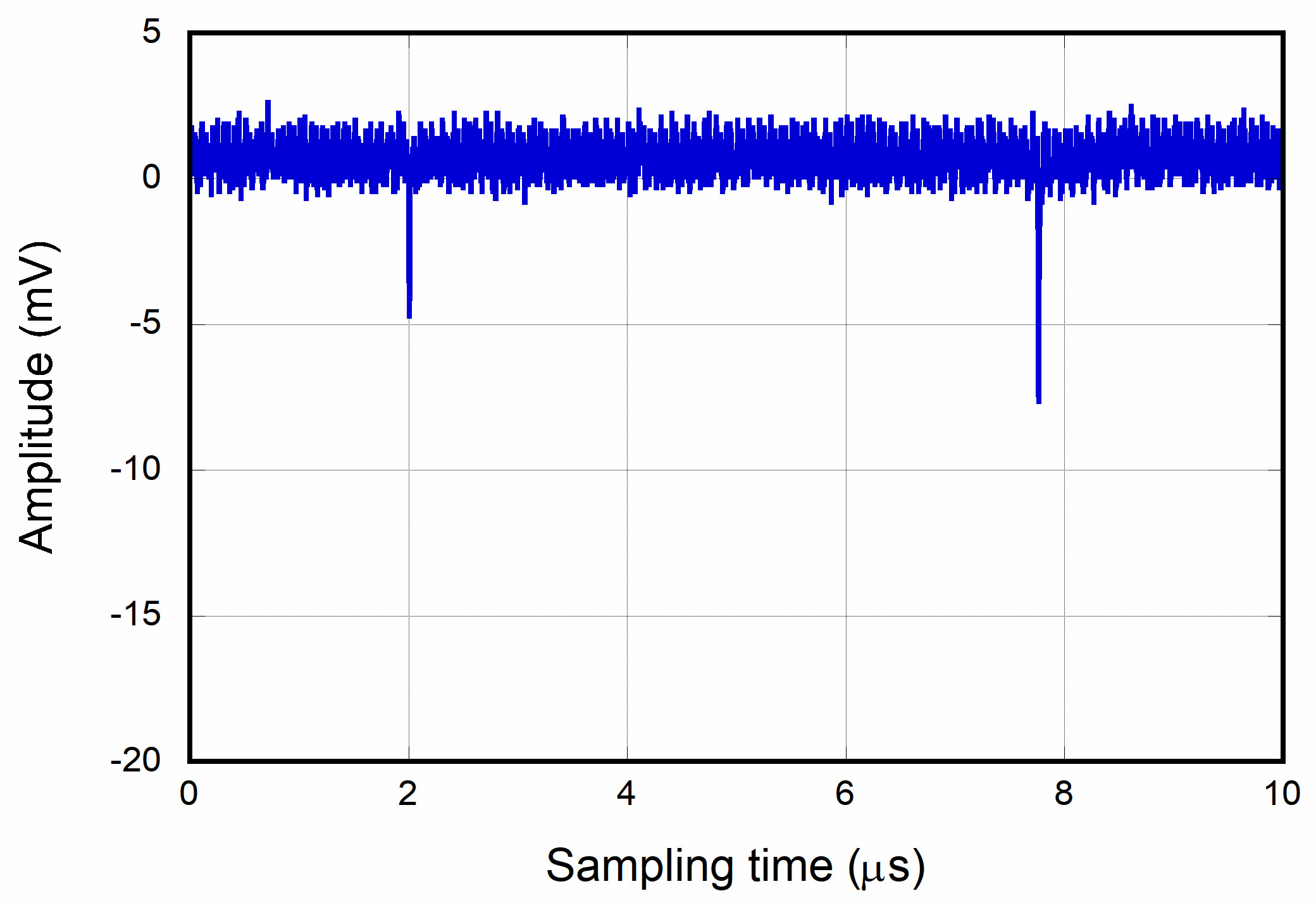} 
\caption{{\it Left}\/: example of actual PMT signal shape recorded in combination with a laser pulse.
{\it Right}\/: a 10~$\mu$s off-light recording showing two random single-electron pulses.}
\label{PMT_signal} 
\end{figure}

The full light detection system was tested at Fermilab after installation, in order to check the
functioning of all the PMT channels, evaluate their performance before
cooling down, %gain knowledge on how to perform calibration at commissioning stage
% gain knowledge on how to perform calibration at commissioning stage
evaluate the effectiveness of the internal optical-fiber
light-distribution system.
To this purpose, %a preliminary set-up was assembled using 
a subset %and instrumentation
of the final electronic chain was used. 
This allowed gaining experience on how to program the
new electronics and how to synchronize it with other detector subsystems, such as the trigger and the TPC
electronics.
%A CAEN V1730B digitizers was used for signal
%recording, allowing the digitization of 16 channels at 500~MHz sampling rate and 14-bit resolution (0.122 mV/bit). 
%This digitizer model %was selected following a test campaign carried out at CERN with a 10-PMT LAr-TPC prototype
%offers the possibility 
%%%During the acquisition, data stream is continuously written in circular memory
%%%buffers %. For each channel the total digital memory is 5.12 MSa, divided into 1024 buffers 
%%%of 10~$\mu$s each. This value is adopted by the experiment to allow, for each event during operation with LAr, the recording of both components of the LAr
%%%scintillation light, i.e. photons from fast and slow decays of exited excimers to ground state.
%%%When a trigger signal occurs, common to all the 16 channels, the active buffer is frozen %, writing operations are moved to the next
%free buffer 
%%%and stored data are ready for read out. %by %the DAQ. 
%Data readout is performed 
%%%Board programming and data read-out is performed by means of an 80~MB/s optical link to a CAEN A3818C housed in the
%%%PC DAQ computer running the DAQ software. 
%%%The PMT voltages were provided by the ICARUS PMT power supply chain, %generation and distribution chain
%%%consisting of a BERTAN 210-02R for the generation a primary -2000~V voltage 
%%%and a CAEN SY1527 crate with A1932AN boards for voltage regulation and distribution.
Test data were recorded with triggers generated by means of a pulse generator with and without the combination of
%both in combination and not with 
laser light pulses %(405~nm, 50~ps) 
through the optical-fiber
light-distribution system.
An example of PMT signal shape recorded
in combination with a laser light pulse and a 10~$\mu$s random trigger acquisition are shown in figure~\ref{PMT_signal}.

%}

A first quick check was carried out to highlight possible problems related to
the detector transfer to Fermilab. 
We found 351 working PMTs, a PMT sparking when applied voltage exceeded
$957$~V, two dead PMTs, and 6 PMTs with some issue when
illuminated with the laser source. These PMTs are are undergoing further investigation.

\begin{figure}
\center
\includegraphics[width=0.49\columnwidth]{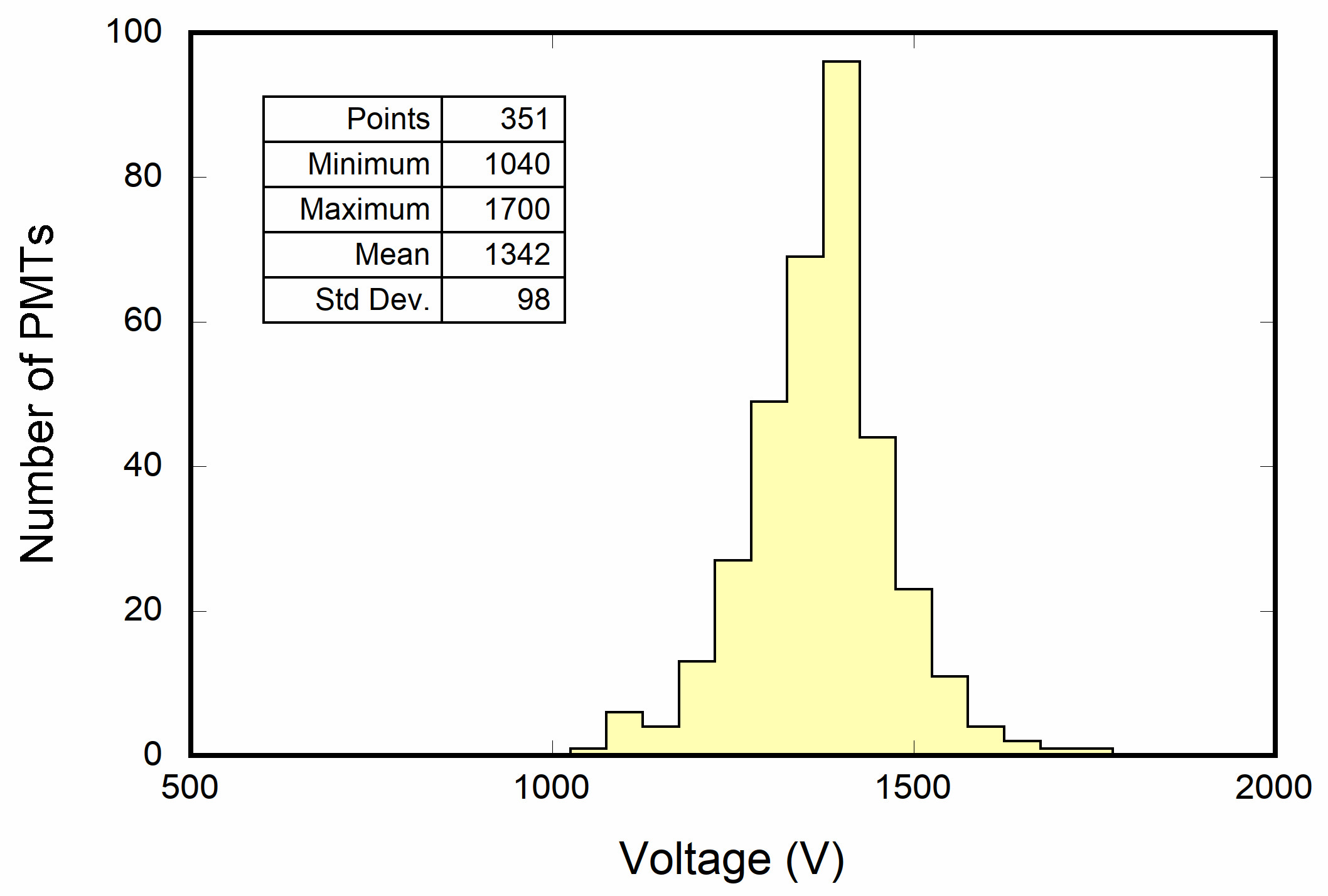} 
\includegraphics[width=0.49\columnwidth]{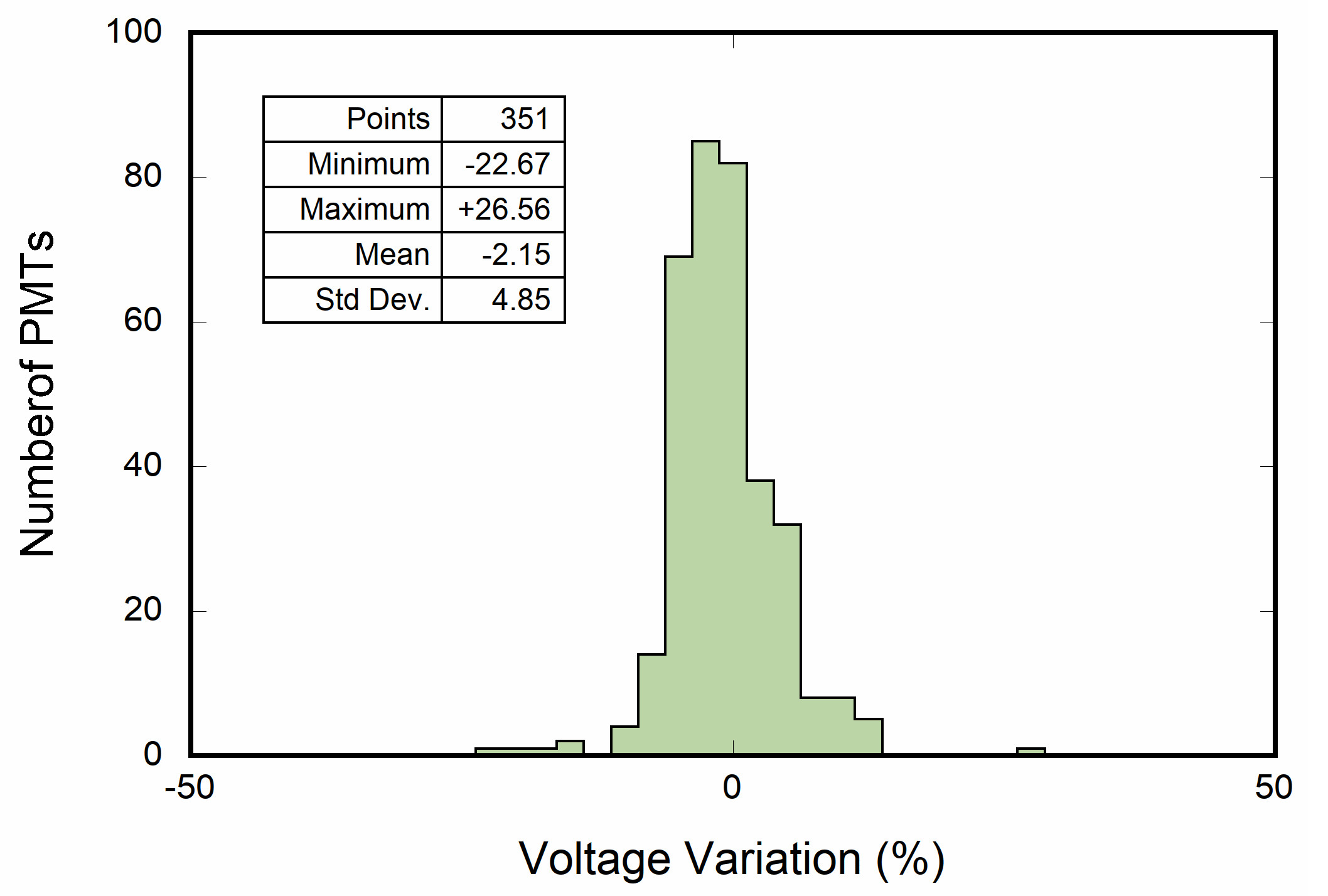} 
\caption{{\it Left}\/: distribution of voltages to attain a gain $G=10^7$ for a set of 351 PMTs. {\it Right}\/: distribution of 
voltage variation (percentage)  with respect the calibration performed at CERN~\cite{Babicz:2018svg}.
The results are consistent with a standard deviation of about 5\%.}
\label{GainAll} 
\end{figure}

The PMT signal analysis was mainly focused on the gain calibration and dark count rate.
The gain calibration of the working PMTs was carried out by acquiring PMT waveforms at a
minimum of 3 voltage points.
%with a Tektronix MSO6 oscilloscope and simultaneously lighting up 10 devices with an external LED (405~nm, 50~ps). 
The gain was evaluated by fitting the charge distribution with the analytical expression
described in~\cite{BELLAMY1994468}. In figure~\ref{GainAll}, 
the distribution of the applied voltage needed to attain a gain of $G=10^7$ for the
351 working PMTs is shown. Results are consistent with a standard deviation of about 5\% 
with respect the calibration performed at CERN~\cite{Babicz:2018svg}.

The dark rate of PMTs %measured using LED off data and applying a random trigger. It 
was evaluated at Fermilab
by counting the number of random single photoelectron pulses 
using light off data %in a 20~$\mu$s windows 
and dividing the total number
by the effective acquisition live time. An overall average rate of about 1.6~kHz was found.
This value is consistent with the 
average rate of 2.1~kHz previously measured at room temperature before the installation of the PMTs,
taking into account that a different measurement technique was adopted at CERN~\cite{Babicz:2018svg}.

\section{Conclusions}

The new scintillation light detection system for the ICARUS T600 LAr-TPC, realized for its
operation at Fermilab in the context of the SBN program, 
includes the use of 360 large
area PMTs mounted behind the wire planes and a fast-laser calibration system.
The high performance of this detection system in terms of sensitivity, granularity and
time resolution, will allow ICARUS to cope with the
large cosmic ray background by identifying the events associated with the neutrino beam.

To this purpose the system was extensively simulated, the components were
precisely characterized and the installation procedures and techniques were carefully defined. 

Preliminary tests carried out after transport and installation of the apparatus at Fermilab verified
the performances of the light detection system required for the identification of signals related
to neutrino beam induced events.

%This system will be completed with 

%A suitable electronics based on fast digitizers
%allows the recording of the waveforms of each PMT signal and provides
%the trigger system with logic pulses.
%which, combined with information
%coming from other detector subsystems, will permit the generation of the
%ICARUS T600 global system trigger signals.
%which will be 
%described in dedicated papers.

\section*{Acknowledgment}

This work was funded by INFN in the framework of the CERN WA104/NP01 program finalized to the
overhauling of ICARUS detector %in view of its operation on SBN at Fermilab.
and was supported by the Fermi National Accelerator Laboratory under
US Department of Energy contract No. DE-AC02-07CH11359.
%the US Department of Energy.
The research activities were also subsidized %supported 
by the EU Horizon 2020 Research and
Innovation Programme under the Marie Sklodowska-Curie Grant
Agreement No. 822185.

%\section*{References}

\bibliography{bibfile}

\end{document}